\documentclass{IEEEtran}
\usepackage{upgreek}
\usepackage{graphicx}
\usepackage{subfigure}
\usepackage{amsmath,bm}
\usepackage{enumerate}
\usepackage[mathscr]{euscript}
\usepackage{color}
\usepackage{url}

\ifCLASSINFOpdf
\else
\fi

\begin{document}

\title{NLOS Transmission Analysis for Mobile SLIPT Using Resonant Beam}

\long\def\symbolfootnote[#1]#2{\begingroup%
\def\thefootnote{\fnsymbol{footnote}}\footnote[#1]{#2}\endgroup}
\renewcommand{\thefootnote}{\fnsymbol{footnote}}
\author{Mingqing~Liu,
Shuaifan Xia,
Mingliang Xiong,
Mengyuan~Xu,\\
Qingwen~Liu,~\IEEEmembership{\normalsize Senior Member,~IEEE\normalsize},
and Hao~Deng

%
\thanks{
Mingqing Liu, Shuaifan Xia, Mingliang Xiong, Mengyuan Xu, and Qingwen Liu are with College of Electronics and Information Engineering, Tongji University, Shanghai 201804, China (e-mail: clare@tongji.edu.cn,
collinxia911@gmail.com,
xiongml@tongji.edu.cn,
xumy@tongji.edu.cn,  qliu@tongji.edu.cn).

Hao Deng is with the School of Software Engineering, Tongji University, Shanghai 201804, China (e-mail: denghao1984@tongji.edu.cn).
}}


\maketitle

\begin{abstract}
Simultaneous lightwave information and power transfer (SLIPT) is a potential way to meet the demands of sustainable power supply and high-rate data transfer in next-generation networks. Although resonant beam-based SLIPT (RB-SLIPT) can realize high-power energy transfer, high-rate data transfer, human safety, and self-alignment simultaneously, mobile transmission channel (MTC) analysis under non-line-of-sight (NLOS) propagation has not been investigated. In this paper, we propose analytical models and simulation tools for reflector-assisted NLOS transmission of RB-SLIPT, where transmission loss and accurate beam field profile of NLOS MTC can be obtained with a receiver at arbitrary positions and attitude angles. We establish analytical models relying on full diffraction theory for beam propagation between tilted or off-axis planes. Then, we provide three numerical methods (i.e., NUFFT-based, cubic interpolation-based, and linear interpolation-based methods) in simulations. Moreover, to deal with the contradiction between limited computing memory and high sampling requirements for long-range transmission analysis, we propose a multi-hop sliding window approach, which can reduce the sampling number by a factor of thousands. Finally, numerical results demonstrate that RB-SLIPT can achieve $4$W charging power and $12$bit/s/Hz data rate over $2$m distance in NLOS scenarios.

\end{abstract}

\IEEEpeerreviewmaketitle

\section{Introduction}
\label{sec:Introduction}
\IEEEPARstart{W}{ith} growing prosperity of applications in smart homes, smart industry, smart cities, etc., the number of Internet of Things (IoT) devices is growing exponentially~\cite{david20186g}. Meanwhile, the high-rate data transfer and constant power supply are increasingly demanding. Utilizing the electromagnetic wave to deliver energy and data simultaneously has been investigated sufficiently~\cite{SWIPT2}. Research on simultaneous lightwave information and power transfer (SLIPT) has gained attention from both academia and industry due to its potential performance and spectrum 
expansion~\cite{OWIPT1,SLIPT3}. Among the existing SLIPT schemes which generally utilize light-emitting diode (LED)~\cite{VLC1,VLC2,VLC3,VLC4,VLC5} or lasers~\cite{LaserSLIPT1,LaserSLIPT2} as the transmitter, a resonant beam-based SLIPT (RB-SLIPT) scheme has been put forward to simultaneously realize high transmission efficiency and self-alignment without beam-steering control~\cite{mobility2}.
However, existing SLIPT schemes, including RB-SLIPT, have a line-of-sight (LOS) transmission channel. At the same time, studying system operation under non-line-of-sight (NLOS) is meaningful for practical implementation in realistic application scenarios. Thus, we investigate the RB-SLIPT under NLOS propagation for the first time, especially emphasizing the mobile transmission channel (MTC) analysis in NLOS scenarios.

\begin{figure}[htbp]
    \centering
     \includegraphics[width=3.5in]{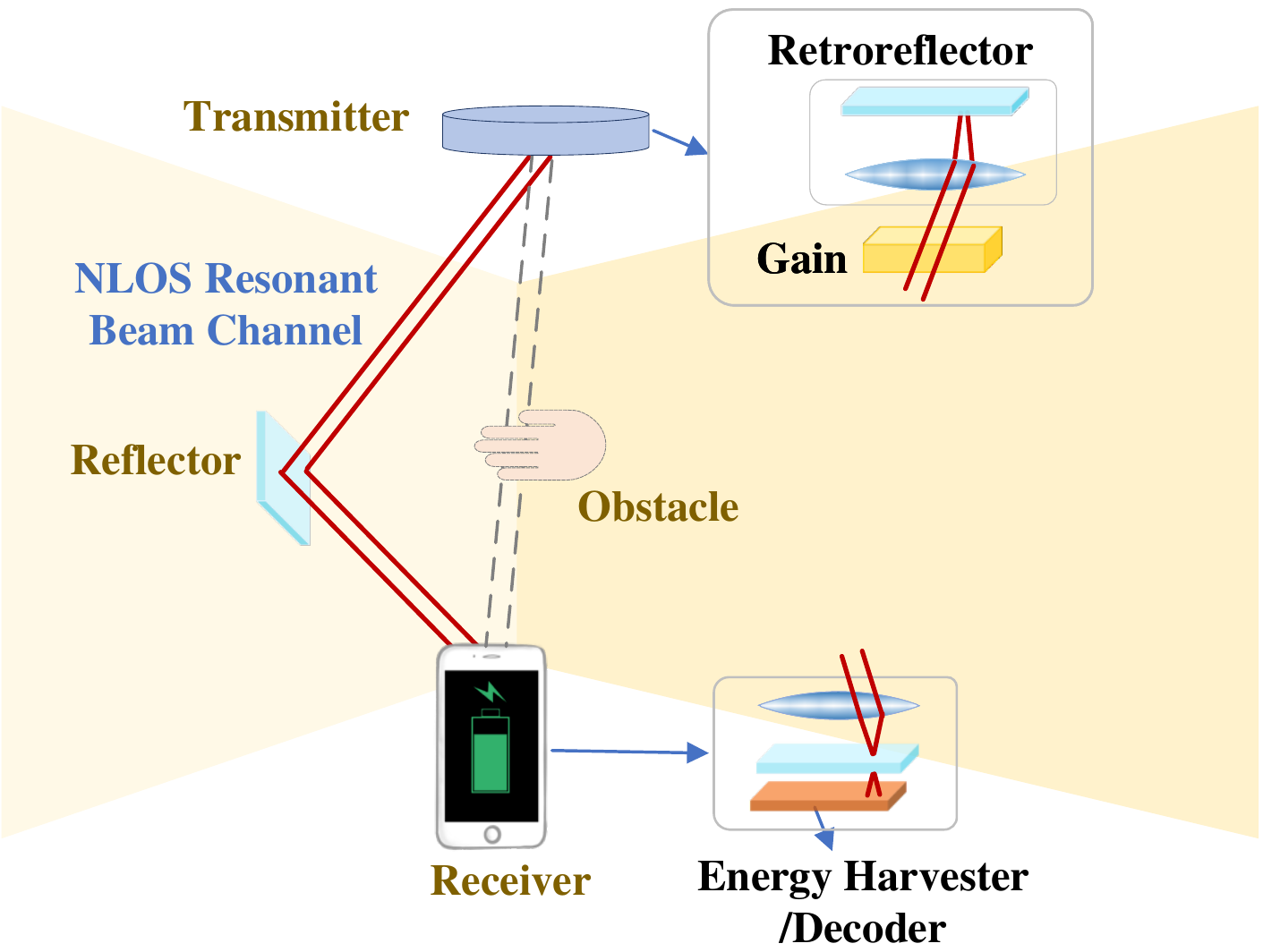}
    \caption{Resonant beam system application for power/data transfer and positioning.}
    \label{f:application}
\end{figure}
Recently, considerable efforts have been put into developing channel models and analysis methods for lightwave-based systems, i.e., optical wireless communications (OWC) and visible light-based positioning (VLP), under NLOS propagation~\cite{VLPNLOS1,VLPNLOS2,NLOSMIMO}. Generally, there are two ways of dealing with NLOS scenarios: i) building models for NLOS channels and analyzing performance limits of systems; ii) treating NLOS links as disturbance sources and will not provide information contribution~\cite{VLPNLOS1}. However, the latter method leads to limited system performance. Thus, research on link performance analysis, channel modeling with multiple-input multiple-output (MIMO) configurations, and channel impulse response analysis have been investigated for OWC in NLOS scenarios~\cite{NLOSMIMO,NLOSCIR,NLOSReflection}. Moreover, a reflection-assisted NLOS channel model has been built in ultraviolet communications, and the reflector can reduce the path loss \cite{NLOSRelay}. Besides, millimeter wave-based system is also susceptible to blockage and suffers higher path loss \cite{millimeter}. Thus, an NLOS channel has been established for enhancing wireless power transfer (WPT) efficiency with the assistance of a reflective surface. Therefore, NLOS channel study is essential for indoor communications/positioning/WPT systems, and the reflection-assisted scheme brings system enhancement. However, NLOS transmission analysis of RB-SLIPT with assistance of the reflector has not been investigated.

RB-SLIPT relies on the structure of the resonant beam system (RBS). RBS consists of a transmitter and a remotely placed receiver, which jointly form a laser resonator~\cite{Liu2016Charging,Xiong2}. The retroreflector embedded in the transceivers can reflect the beam back along its incoming direction. Thus, beams can be reflected back and forth within the laser resonator without the limitation that the transceiver must be strictly faced with each other. With sufficient input power, the intra-cavity laser (i.e., named as resonant beam) can be self-established as the receiver is placed within the transmitter's field of view (FoV). That's the principle of self-alignment nature~\cite{mobility1,mobility2}. Besides, the resonant beam will be cut off immediately as the foreign object is close to the transmission channel, guaranteeing safety with larger deliverable power~\cite{RBCSafety}. 

With the above features, the data/energy transfer with RB-SLIPT terminates if there is an obstacle, e.g., a human hand, in the LOS link between the transceiver as in Fig.\ref{f:application}. However, suppose there is a reflector placed nearby the LOS link. In that case, a transmission channel can be established through the RBS transmitter, reflector, and RBS receiver, where data and energy can be transferred. According to the reflection principle, the NLOS transmission channel is equivalent to the transmission channel between a pair of non-coaxial transmitters and receivers. Thus, to analyze the RB-SLIPT performance with a receiver placed at an arbitrary position and attitude angle under NLoS propagation, we can conduct the RB-SLIPT analysis with an arbitrarily off-axis and tilted receiver under LOS propagation, i.e., mobility analysis of RB-SLIPT.

The mobility mechanism of RB-SLIPT has been revealed theoretically, and the system performance, i.e., data rate, charging power, and positioning accuracy within a specific FoV, can be investigated with the analytical models~\cite{mobility1,mobility2}. Given the parameters of the gain medium and the resonant cavity, such as retroreflector size, we should at first obtain the transmission loss (i.e., diffraction loss during the long-range beam transmission within the cavity) of the resonant beam. The output laser power of the laser resonator can be calculated with the methods for traditional lasers. Finally, the SLIPT performance can be analyzed using the energy/information separation schemes in the receiver design. Moreover, RBS has the potential for passive positioning \cite{positioning}, where the beam field profile on the mirror in the transmitter is required. Above all, building the analytical model for MTC of RB-SLIPT to obtain transmission loss and accurate beam profile (i.e., self-reproducing mode) is necessary. 

For building the MTC model with a moving receiver and two retroreflectors, an equivalent resonator model is proposed to prove that the double-retroreflector cavity with one 
non-coaxial retroreflector can be equivalent to a Fabry–Pérot (FP) cavity with reduced reflection area due to the mobility~\cite{mobility1,mobility2}. However, to obtain an accurate beam profile, including position information of the resonant beam, a method that utilizes the beam transmission between non-coaxial planes or tilted lanes should be developed for transmission channel analysis under NLOS propagation of RB-SLIPT. The beam transmission through each element in the cavity is simulated to direct the calculation the self-reproducing mode without the equivalence. This manuscript focuses on the MTC analysis with an off-axis and titled receiver to illustrate the RB-SLIPT performance in NLOS scenarios. We also developed a complete implementation method and simulation model for RB-SLIPT mobility analysis with arbitrarily displaced receivers.

The contributions of this manuscript are:

\begin{itemize}
\item[C1)] We propose analytical models for reflector-assisted NLOS transmission of RB-SLIPT, where the transmission loss and accurate beam field profile of MTC can be obtained. Based on the above models, we can analyze the energy/data transfer performance of RB-SLIPT with a receiver at arbitrary positions and attitude angles in the NLOS scenarios.

\item[C2)] We develop a computational approach with multi-hop sliding windows for long-range transmission simulation, which can reduce the sampling number by a factor of thousands under $2$m distance compared to the original methods. 
\end{itemize}

The remainder of this paper is organized as follows. In Section II, we describe the system architecture and formulate the problem of RB-SLIPT MTC analysis under NLOS propagation. In Section III, we present the analytical models, including the self-reproducing mode and transmission loss calculation, in which the receiver can be arbitrarily located, i.e., non-coaxial and/or tilted with respect to the transmitter. In Section IV, we present the implementation details during the computational process. In Section V, we demonstrate the RB-SLIPT performance with different moving states under NLOS propagation through numerical analysis. Finally, we make a conclusion in Section VI.

\section{Problem Formulation}
\label{sec:SysOverview}
\begin{figure}[!t]
    \centering
     \includegraphics[width=3.5in]{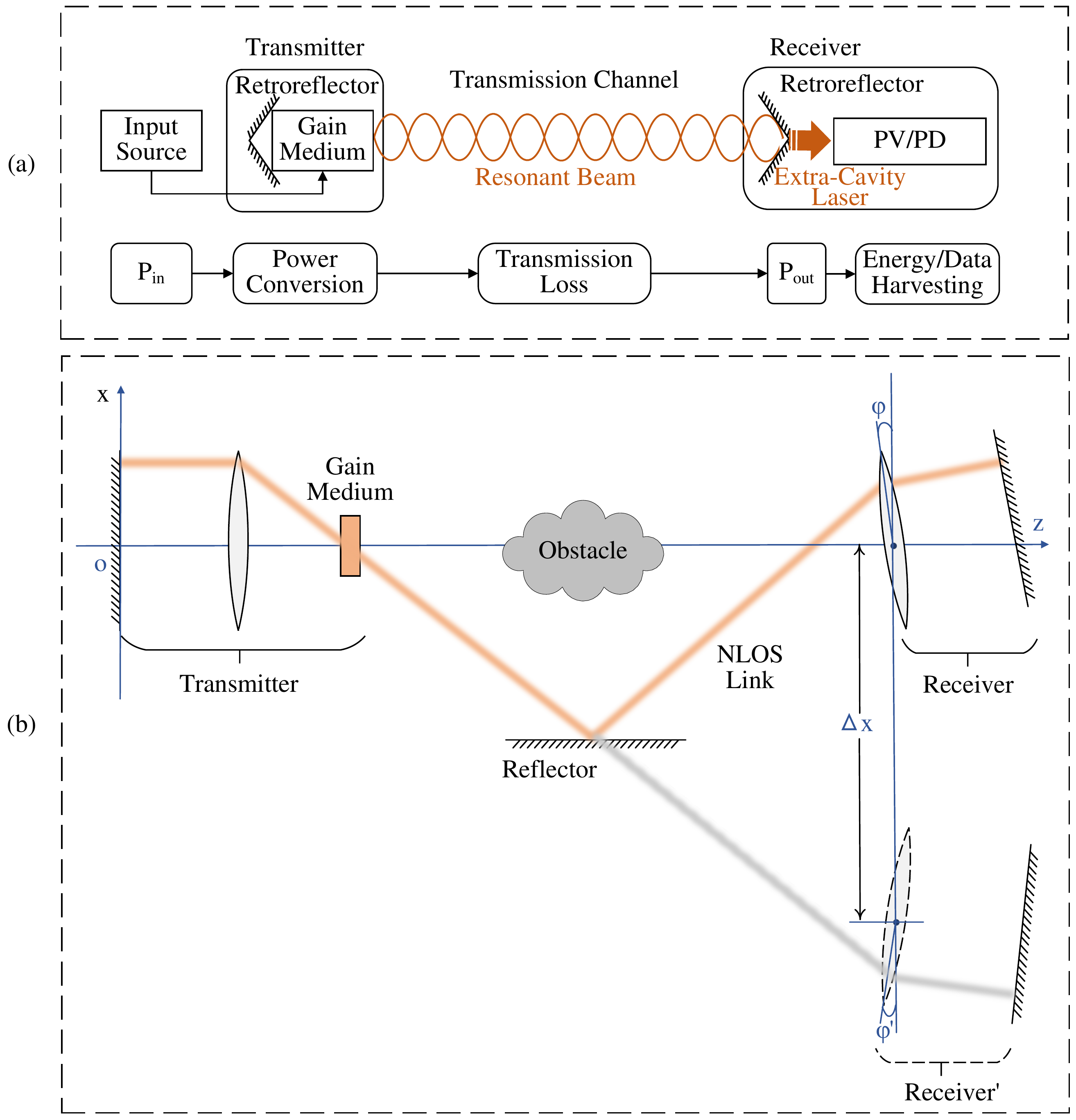}
    \caption{(a) Structure and (b) mobility illustration of resonant beam system.}
    \label{f:sysandprob}
\end{figure}

\subsection{RB-SLIPT Structure and Feature}
In Fig.~\ref{f:sysandprob}(a), RB-SLIPT consists of a spatially separated transmitter and receiver, forming a resonant cavity. The resonant beam generated within the resonant cavity (i.e., transmission channel) is essentially an intra-cavity laser. Resonant beam acts as a carrier of energy or data transferring over-the-air. We can depict the generation of the resonant beam by describing the laser generation principle: there is an input source, a gain medium, and a resonant cavity formed by two mirrors which allows the beam to be reflected back and forth; the input source stimulates the gain medium to transmit photons, and photons bounced within the cavity can pass the gain medium multiple times to be amplified; once the amplification the photons experienced can compensate for the total losses within the cavity, a stable laser beam can be established. The laser beam in the cavity is called a resonant beam, and if one of the mirrors of the resonant cavity is transmissive, a part of the beam passes through it, yielding the 
extra-cavity laser beam. After power splitting, one part of the laser beam can be converted by a photovoltaic (PV) panel to electrical power and ready for charging; the other part is received by the photodiode (PD) for data collection. The power flow in RBS is also depicted in Fig.~\ref{f:sysandprob}(a). $P_{\rm in}$ from the input source is given to the gain medium, and the electrical power is converted to beam power. Then, after transferring over-the-air with transmission losses, the laser beam is output with power $P_{\rm out}$. Finally, the electrical power converted by PV and data rate from PD can be analyzed. According to~\cite{iteration,modecalculation,FFT,FoxLi}, one of the most critical parts of the above system model is obtaining accurate transmission loss of the RB-SLIPT channel.

The employment of resonant beam brings two advantages: self-alignment and safety. If we take retroreflectors (an optical element that can reflect beams from any direction back along the incoming path) as mirrors to form the resonant cavity, the resonant beam can be generated even if the two retroreflectors are not strictly faced with each other. If the receiver with retroreflector is within the FoV of the transmitter, the transmission channel can be self-established between the transceivers, leading to the self-alignment feature. Moreover, if there is a foreign object (i.e., human hands) invading the transmission channel, the generation condition of the resonant beam will be destroyed, and the resonant beam ceases immediately. Thus, the high-power beam will not hurt humans, guaranteeing human safety. 
 
 \subsection{NLOS propagation}
 RB-SLIPT utilizes lightwave as the carrier of both information and power, which is sensitive to blockage. If a foreign object is placed within the transmission channel between the transceiver, the resonant beam transmission will be cut off, and SLIPT will be terminated. Here we review the three conditions of resonant beam generation in RBS: i) pump source, ii) gain medium, and iii) resonant cavity that allows the beam to be reflected back and forth multiple times 
within it. Under NLOS propagation, the former two conditions remain. If we can construct a loop so that the resonant beam can still travel back and forth in the resonant cavity and through the gain medium after being reflected by an additional reflector, the resonant beam can be generated. As in Fig.~\ref{f:sysandprob}(b), an obstacle blocks the transmission channel between the transmitter and receiver. If we attach a reflector, i.e., a mirror, to the wall beneath the transmission channel, a resonant beam can be established between the transmitter, reflector, and receiver.

According to the mirror reflection principle, the beam path from the transmitter to the reflector and then the receiver is equivalent to the beam path from the transmitter to the receiver' neglecting the losses at the mirror, and receiver and receiver' are mirror-symmetrical. Thus, we build the coordinate system to illustrate the NLOS propagation of RBS as in Fig.~\ref{f:sysandprob}(b). We take the resonant cavity with two cat's eye retroreflectors as an example. Cat's eye consists of a lens with focal length $f$ and a mirror of which the radius is both $r$. The origin of the coordinate system is consistent with the center of the mirror of the cat's eye in the transmitter, and the x-axis coincides with its surface. Moreover, the z-axis is perpendicular to the cat's eye in the transmitter. Suppose the two cat's eyes are strictly aligned to each other, and the distance along the z-axis between the transceivers is $\Delta z$. If the reflector is placed at $z=\Delta z/2$, the vertical distance of the reflector to the z-axis $h$ needs to satisfy
\begin{equation}
    h\le r_c\Delta z/2f.
\end{equation}
 Then, the beam path under NLOS propagation is equivalent to the receiver's position on the z-axis by $\Delta x=2d$ on the x-axis.

 Moreover, the receiver of the RB-SLIPT mat does not face the transmitter strictly. As in Fig.~\ref{f:sysandprob}(b), the receiver is coaxial with the transmitter but tilts $\varphi^{\circ}$ on the y-axis. Correspondingly, the tilt angle of the mirror-symmetrical receiver' is $\varphi'=-\varphi$. Above all, the transmission channel analysis of RB-SLIPT under NLOS propagation can be deduced to the analysis that the receiver is with the two moving states: i) tilts $\varphi'^{\circ}$ on the x-axis and ii) off the z-axis by $\Delta x$m along x-axis depending on the position the reflector is placed. It can be deduced that as the $\varphi^{\circ}$ and $\Delta x$ grow, the output power from the system reduces. Define that RB-SLIPT is workable as $P_{\rm out}>0$, we can obtain the maximum allowable moving range and tilt angle of the receiver in RB-SLIPT by calculation of $P_{\rm out}$.
  
 Thus, we should obtain $P_{\rm out}$ with different reflector positions and receiver's moving along the z-axis or tilt angles, where the key is to get the transmission loss as the power conversion process can be modeled as analytical models. To obtain the transmission loss, we adopt the electromagnetic field propagation method to simulate the beam transmission process in the system as the receiver is off-axis or tilted. Thus, the problem can be formulated as finding the transmission loss with different $\Delta x $, $ \Delta y$, $\Delta z$, and $\varphi$.
 
 \begin{figure}[h]
    \centering
    \includegraphics[width=3.5in]{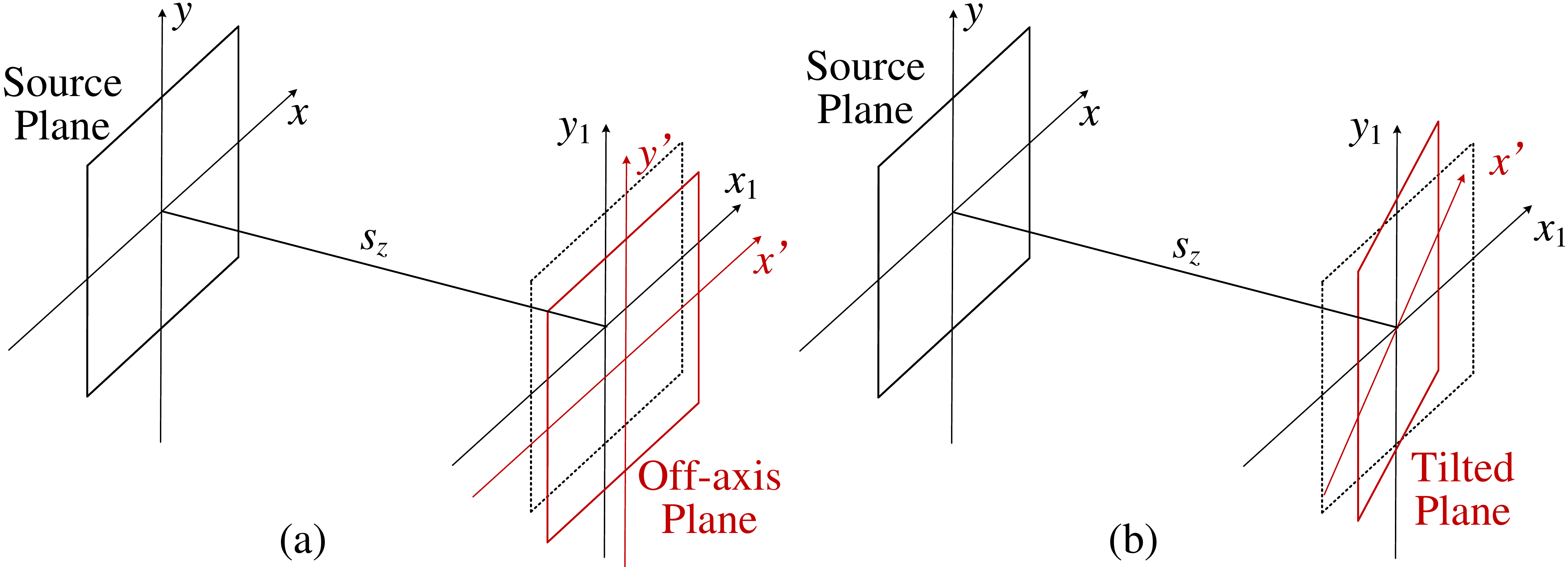}
    \caption{Coordinate illustration for (a) off-axis and (b) tilted planes.}
    \label{f:coord-def}
\end{figure}

 \begin{figure}[h]
    \centering
    \includegraphics[width=3.3in]{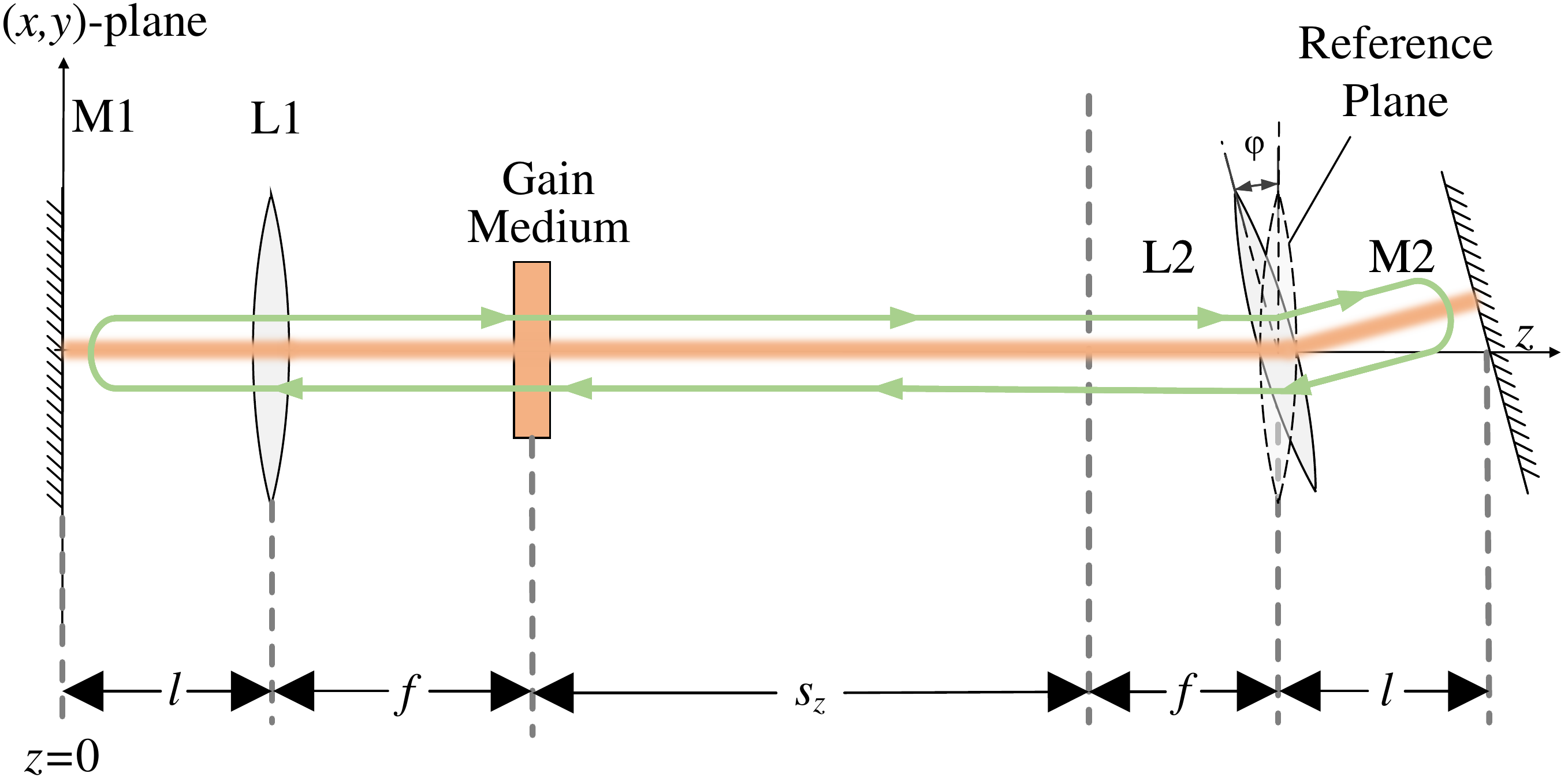}
    \caption{Illustration of RBS with a tilted receiver.}
    \label{f:tilted-arch}
\end{figure}
 
\section{Mobile Transmission Channel Modelling}
\label{sec:PositionModel}
Based on the Rayleigh-Sommerfeld (RS) diffraction integral, the spatial transfer of an optical beam field can be simulated without using Fresnel diffraction approximation~\cite{FFT}. The only approximation is to neglect the vectorial nature of light, which can be compensated with the RS integral. Besides, the simulation for beam field can be extended to some particular circumstances like beam transfer between non-coaxial planes and tilted planes as shown in Fig.~\ref{f:coord-def} (a) and (b) respectively.

To conduct the performance analysis for RB-SLIPT under NLOS propagation, we should at first obtain the self-reproducing mode of the resonant beam in the resonant cavity with an off-axis and tilted cat's eye retroreflector at the receiver. Then, the diffraction loss, which we regard as the over-the-air transmission loss, can be calculated, and the energy/data transfer performance under moving status can be evaluated. To obtain the self-reproducing mode, we should build the self-consistent equation for round-trip transmission of the resonant beam by simulating the beam field propagating through each optical element and free space in the RBS. Obviously, as the receiver is with an arbitrary attitude angle or moves to an arbitrary spatial position within the system's FoV, the crucial part of the simulation is the free space between the transceivers, as shown in Fig.~\ref{f:coord-def}, where the diffraction between off-axis or tilted planes can be calculated with RS diffraction integral. This section gives the establishment of a self-consistent equation with moved receivers and explicates the method for performance analysis under mobility.

\subsection{Receiver with Arbitrary Attitude Angle}
We at first build the analysis model for a receiver of RBS with an arbitrary attitude angle as in Fig.~\ref{f:tilted-arch}. The diffraction transmission between tilted planes is depicted, after which we establish the self-consistent equation for simulating the round-trip transmission of the resonant beam with a tilted receiver.
\subsubsection{Diffraction Transmission between Tilted Planes}
Fig.~\ref{f:diff-tilt} illustrates the situation where the receiver's plane is tilted as we take the plane the gain medium is placed as the source plane, and the plane the lens L2 in the receiver as the tilted plane. The tilted plane is tilted $\varphi^{\circ}$ relative to and $s_z$ distance away from the source plane. Simulation of beam transfer between tilted planes can be summarized in the following two steps~\cite{shift,tiltMost}:
\begin{figure}[h]
    \centering
    \includegraphics[width=2.5in]{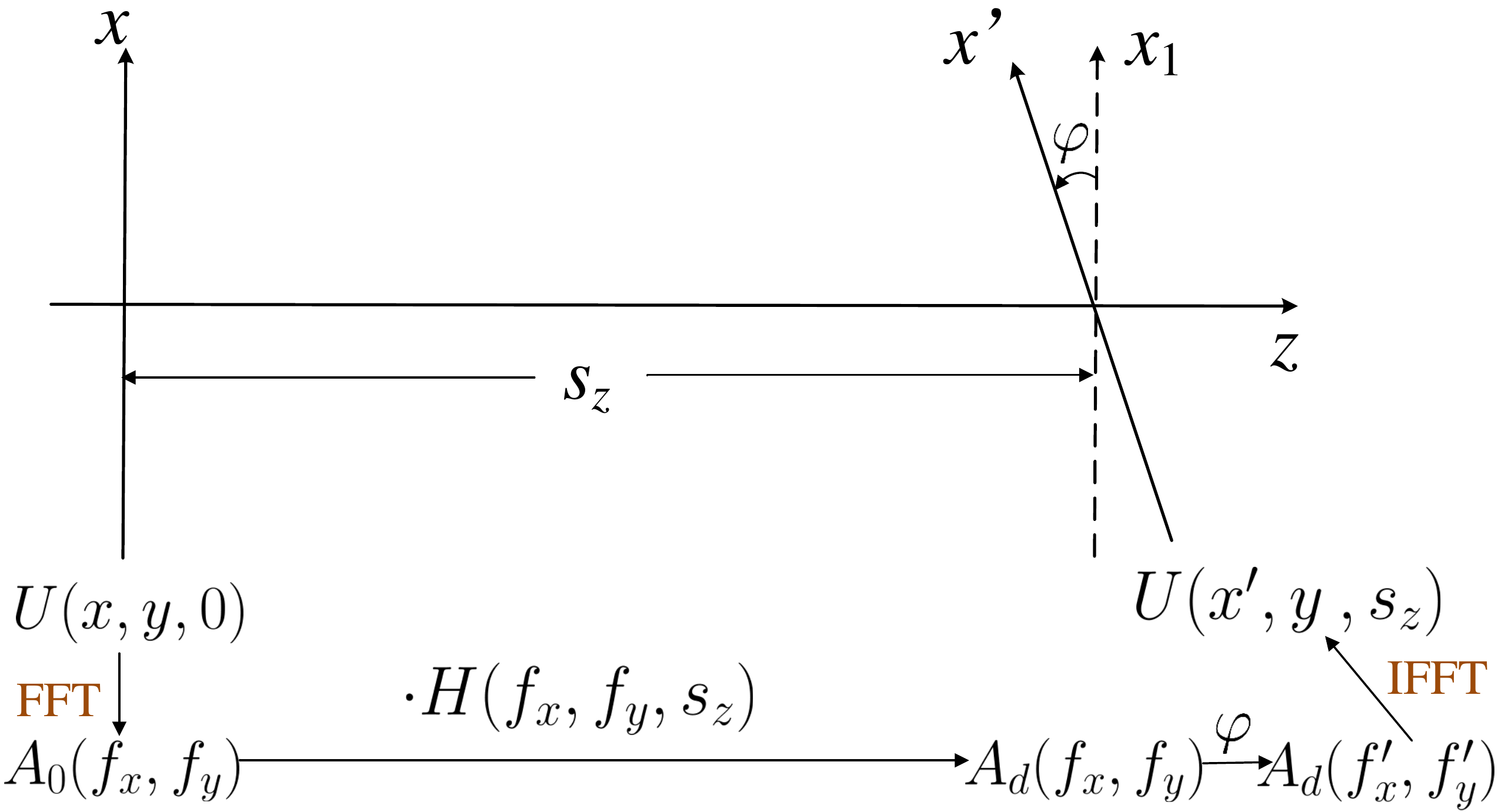}
    \caption{Diffraction transfer between tilted planes.}
    \label{f:diff-tilt}
\end{figure}

\begin{itemize}
\item[i)]The beam field first propagates to a reference plane $(x_1, y')$ with the same origin as the tilted plane but is parallel to the transmitter's plane. This propagation process can be simulated exactly the same as the diffraction between parallel planes. Namely, we obtain the angular spectrum of the original beam field distribution by applying Fourier transform to it and, in the frequency domain, multiplying with the free space transfer function $H(f_x, f_y, s_z)$ parameterized by the z-axis distance $s_z$. This step yields the spectral optical field distribution $A_d(f_x, f_y)$ on the reference plane; that is~\cite{FFT}
\begin{equation}
A_d(f_x, f_y)   =  \mathcal{F} \left\{U(x, y, 0)\right\}\cdot H(f_x, f_y, s_z), 
\label{e:Ad}
\end{equation}
where $\mathcal{F}$ indicates the fast Fourier transform (FFT) process and $H(f_x, f_y, s_z)$ is derived as~\cite{FFT}
\begin{equation}
H(f_x, f_y, s_z) = \exp{\left[ i\frac{2\pi s_z}{\lambda}  \sqrt{1-(\lambda f_x)^2-(\lambda f_y)^2}\right]}.
\label{e:freespace}
\end{equation}

\item[ii)]We map $A_d(f_x, f_y)$ on the reference plane with coordinate conversion to obtain the angular spectrum of distribution $A'_d(f'_x, f'_y)$ with respect to the tilted plane. Finally, we can calculate the beam field distribution on the surface of L2 at the tilted receiver with an inverse fast Fourier transform (IFFT) on $A'_d(f'_x, f'_y)$; that is
\begin{equation}
U(x', y, s_z) = \mathcal{F}^{-1} \left\{A'_d(f'_x, f'_y)\right\},
\label{e:Ud}
\end{equation}
where $\mathcal{F}^{-1}$ indicates the IFFT process.
\end{itemize}

The key step for the above procedure is the coordinate rotation of $A_d(f_x, f_y)$ to obtain $A'_d(f'_x, f'_y)$. The beam field on the reference plane can be perceived as a superimposition of infinite plane waves. For each frequential coordinate $(f_x, f_y)$, $A_d(f_x, f_y)$ denotes the complex amplitude of one plane wave, of which the wave vector can be represented as
\begin{equation}
\begin{aligned}
    &\bm{k}=2\pi\left[f_x\ f_y\  w(f_x, f_y)\right] \\
    &\quad w(f_x, f_y) = \sqrt{\lambda^{-2}-f_x^2-f_y^2}
\end{aligned}.
\label{e:wave-vector}
\end{equation}
Similarly, the vector on the tilted plane can be written as
\begin{equation}
\begin{aligned}
    \bm{k}'=2\pi\left[f_x'\ f_y'\  w'(f_x', f_y')\right], 
\end{aligned}
\label{e:wave-vector-tilt}
\end{equation}
Suppose $\mathbf{T}$ is the conversion matrix that maps the coordinate from the reference plane onto the tilted plane, and we can obtain the relationship between the above two vectors as

\begin{equation}
\bm{
k' =  }\mathbf T \bm{k,\quad k = } \mathbf T^{-1}\bm k'.
\label{e:coord-map}
\end{equation}

Specifically, in the situation illustrated in Fig.~\ref{f:diff-tilt} where the tilted plane has an angle of $\varphi$ relative to y-axis, the inverse  conversion matrix is:
\begin{equation}
\mathbf T^{-1} = \left[
\begin{array}{ccc}
\cos\varphi &0& \sin\varphi  \\
0 &1& 0 \\
-\sin\varphi&0& \cos\varphi 
\end{array}
\right]
\label{e:Tmatrix}
\end{equation}
Then we can obtain the connection between these two coordinate systems as
\begin{equation}
\begin{aligned}
&f_x = \alpha(f'_x, f'_y) = \cos\varphi f'_x + 0\times f'_y + \sin\varphi w'(f_x', f_y')\\
&f_y = \beta(f'_x, f'_y) = 0\times f'_x + 1\times f'_y + 0\times w'(f_x', f_y')
\end{aligned}
\label{e:fxfy}
\end{equation}
Thus, the relation between two spectra can be deduced as
\begin{equation}
\begin{aligned}
A'_d(f'_x, f'_y) &= A_d(\alpha(f'_x, f'_y), \beta (f'_x, f'_y))\\
&= A_d(\cos\varphi f'_x + \sin\varphi w'(f_x', f_y'), f'_y)
\end{aligned}
\label{e:spectra-relation}
\end{equation}

However, since the transform in Eq.~\ref{e:fxfy} is not linear, the spectrum after the rotation will become non-uniform sampled. This non-uniform sample imposed by the rotation will further lead to energy loss, causing incorrect results after simple IFFT. Two approaches are adopted to solve this issue:

\begin{itemize}
\item[i)] Perform Type-I non-uniform Fourier transform onto $A'_d(f'_x, f'_y)$ to transform a non-uniform spectrum input to a uniform sampled beam field distribution as~\cite{nonuniform1,nonuniform2}
\begin{equation}
\begin{aligned}
U(x', y, s_z) = {\rm NUFFT_1}\left\{ A'_d(f'_x, f'_y)\right\},
\end{aligned}
\label{e:nufft}  
\end{equation}
where ${\rm NUFFT_1}$ indicates the Type-I non-uniform Fourier transform process.
\item[ii)]Apply interpolation onto the non-uniform $A'_d(f'_x, f'_y)$ to obtain the uniform-sampled spectrum distribution, and compensate the energy loss by multiplying with a energy loss function $J(f'_x, f'_y)$. Then, employ the ordinary IFFT to obtain uniform beam field on the tilted plane as \cite{tiltMost}
\begin{equation}
\begin{aligned}
U(x', y, s_z) &= \mathcal{F}^{-1}\left\{ A'_d(f'_x, f'_y)\times J(f'_x, f'_y) \right\}\\& \times \exp{\left[i2\pi f'_{x0}x\right]}
\\
J(f'_x, f'_y) &= \mid \cos\varphi - \frac{f'_x}{w'(f'_x, f'_y)}\sin\varphi\mid
\end{aligned},
\label{e:interpol}
\end{equation}
where the term $\exp{\left[i2\pi f'_{x0}x\right]}$ represents the process of shifting the center frequency $f'_{x0}$ back to zero by constructing a new coordinate system $\hat{f}_x = f'_x - f'_{x0}$. That's because the coordinate conversion in Eq.~\ref{e:spectra-relation} will lead to a central frequency displacement on the spectrum, which poses problem for fast Fourier transform as the frequency center is far from zero. 
\end{itemize}

\subsubsection{Self-consistent equation for round-trip transmission}As in Fig.~\ref{f:tilted-arch}, we should simulate beam field passing through each optical element in the system to build the self-consistent equation for round-trip transmission. The beam field starts from M1, i.e., mirror in the cat's eye at the transmitter, passes through lenses, gain medium, free space, and cat's eye in the receiver, and then transfers back along the path from receiver to the transmitter, and finally arrives at M1 again~\cite{coauthor}. We at first denote $\mathcal{P}$ as the propagation operator which is the solution of Eqs.~\eqref{e:Ad}, ~\eqref{e:spectra-relation}, and~\eqref{e:nufft} or \eqref{e:interpol}:
\begin{equation}
    U(x',y,s_z) = \mathcal{P}[U(x,y,0)].
\end{equation}
For the trip from the transmitter to the receiver, as in Fig.~\ref{f:tilted-arch}, given the field in the plane of the gain medium $U_{\rm G}(x,y,f+l)$, the field in plane of L2 can be derived as $U_{\rm L2}(x,y,s_z+2f+l) = \mathcal{P}[U_{\rm G}(x,y,f+l)]$, where $f$ is the focal length of L1/L2, and $l$ is the interval between M1/M2 and L1/L2. However, for the opposite transfer direction, instead of rotating the coordinate system, \begin{figure}[h]
    \centering
    \includegraphics[width=3.5in]{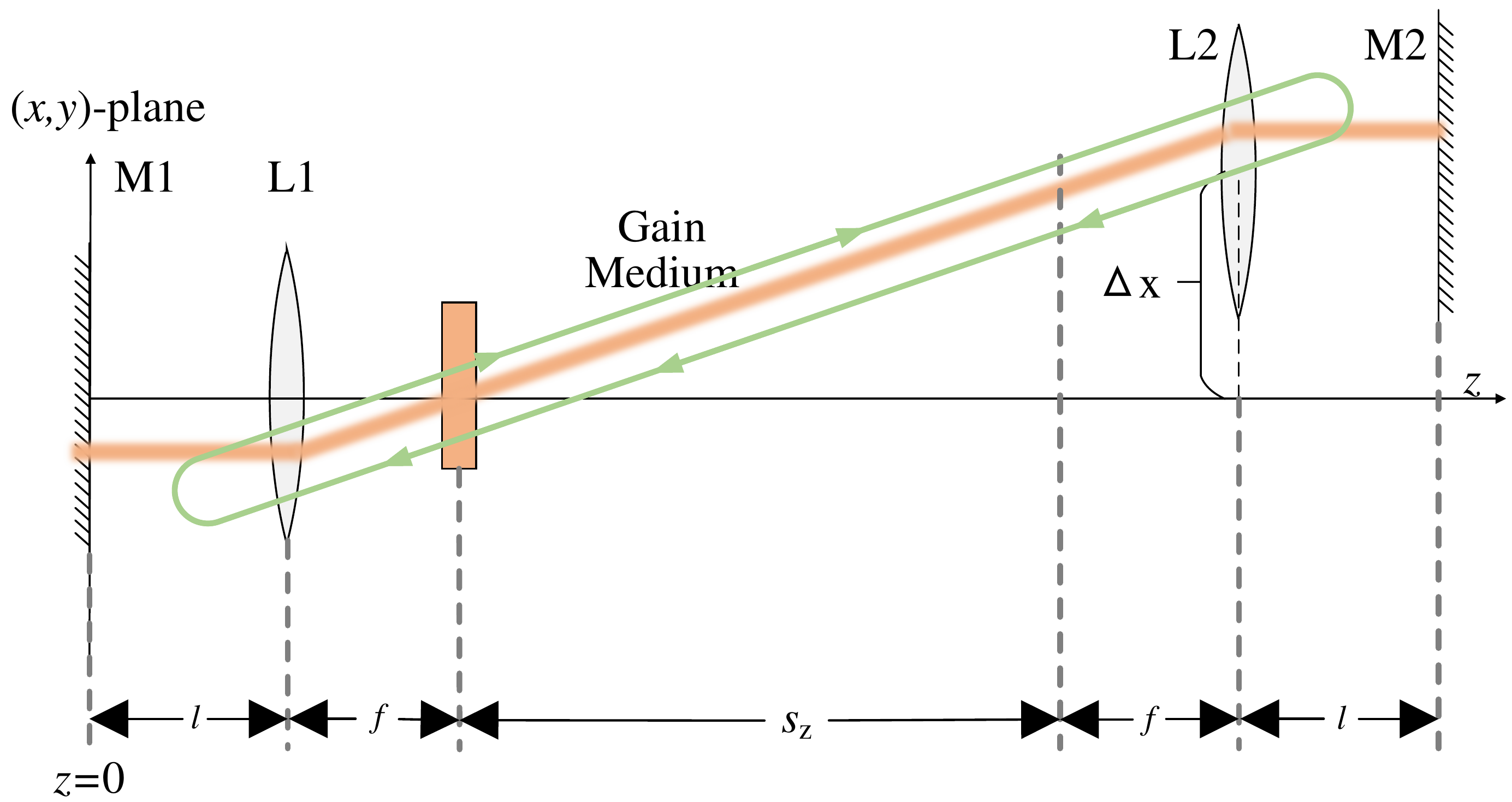}
    \caption{Illustration of RBS with an off-axis receiver.}
    \label{f:offaxis-arch}
\end{figure}We  consider the diffraction transmission is from a tilted plane. The process can be denoted by the propagation operator $\mathcal{P}'$. Similarly, there are two methods to conduct the simulation: 
\begin{itemize}
\item[i)] Perform Type-I non-uniform Fourier transform onto $U_{\rm L2}$ to transform a uniform sampled beam field input to a non-uniform spectrum distribution, and then map the obtained spectrum distribution to the reference plane parallel to plane of the gain medium as Eq.~\eqref{e:spectra-relation} with $-\varphi$. Then, multiply with the free space transfer  function $H(f_x,f_y,s_z+f)$ and obtain $U_{\rm G}(x,y,f+l)$ with IFFT.

\item[ii)] Perform FFT onto $U_{\rm L2}$ and conduct coordinate mapping with $-\varphi$ by applying interpolation onto the non-uniform spectrum distribution to obtain the uniform-sampled spectrum distribution. Then, after multiplying with  $H(f_x,f_y,s_z+f)$, employ the ordinary IFFT to obtain $U_{\rm G}(x,y,f+l)$ on the gain medium as Eq.~\eqref{e:interpol}.
\end{itemize}

Then, we define the propagation operators $\mathcal{A,B,C,D}$ indicating beam transmission through gain medium, lens, free space between mirror and lens, and free space between the lens and gain medium, respectively. The operators are specified with $r_g$ and $r_c$ denoting the radius of the gain medium and cat's eye respectively as follows~\cite{coauthor}:
\begin{equation}
\begin{aligned}
&\mathcal{A}[U(x,y)] = U(x,y)\cdot G(x,y) \\
&G(x, y) = \left\{
\begin{aligned}
&1, x^2 + y^2 \leq r^2_g\\
&0,  else
\end{aligned}
\right.
\end{aligned}
\label{e:gainmedium}
\end{equation}

\begin{equation}
\begin{aligned}
&\mathcal{B}[U(x,y)] = U(x,y)\cdot L(x,y) \\
&L(x, y) = \left\{
\begin{aligned}
&\exp{\left[-i\frac{\pi}{\lambda f}(x^2 + y^2)\right]}, x^2 + y^2 \leq r^2_c\\
&0,  else
\end{aligned}
\right.
\end{aligned}
\label{e:lens}
\end{equation}

\begin{equation}
\mathcal{C}[U(x,y)] = \mathcal{F}^{-1} \left\{ \mathcal{F} \left\{U(x, y)\right\}\cdot H(f_x, f_y, l)\right\}.
\end{equation}

\begin{equation}
\mathcal{D}[U(x,y)] = \mathcal{F}^{-1} \left\{ \mathcal{F} \left\{U(x, y)\right\}\cdot H(f_x, f_y, f)\right\}.
\end{equation}

Thus, taking M1 as the starting plane, we can build the self-consistent equation for round-trip transmission of the resonant beam in RBS where the receiver is with arbitrary attitude as
\begin{equation}
    U(x,y,0) = \mathcal{CBDAP'BCCBPADBC}[U(x,y,0)].
    \label{e:self-consistent}
\end{equation}

\begin{figure}[h]
    \centering
    \includegraphics[width=2in]{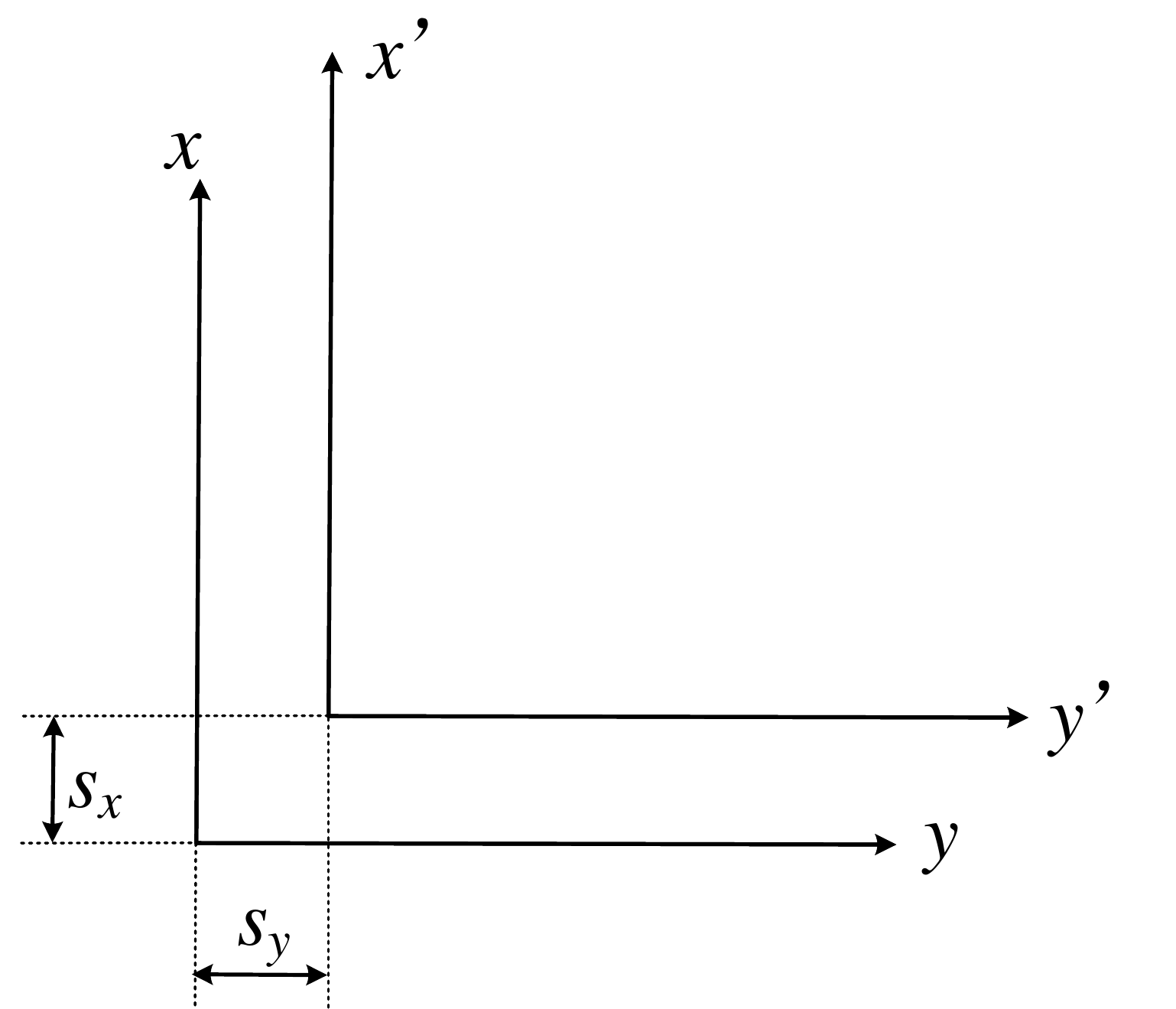}
    \caption{Coordinate shift of off-axis plane.}
    \label{f:coord-shift}
\end{figure}

\subsection{Receiver at Arbitrary Spatial Position}
As in Fig.~\ref{f:offaxis-arch}, the receiver may move to an arbitrary spatial position within the system's FoV, leading to the off-axis of the cat's eye in the receiver relative to the transmitter. We consider a situation that the receiver moves away from z-axis along x-axis with a distance $\Delta x$ and away from the origin along z-axis with a distance $s_z$. Stated that the simulations for receiver moving along y-axis or both x-axis and y-axis are in the same manner.
\subsubsection{Diffraction Transmission between off-axis planes}
The angular spectrum method shed further insight on accelerating the calculation of optical field transmission by applying FFT on the basis of RS diffraction integral. As in Fig.~\ref{f:coord-shift}, the coordinate system of the off-axis plane can be interpreted as a shift from the original coordinate, namely~\cite{shift}:
\begin{equation}
\begin{aligned}
 x' = x - s_x, 
 y' = y-s_y, 
 z' = s_z 
\end{aligned}
\label{e:coord_shift}
\end{equation}
Then, the spatial distribution on the off-axis plane coordinate can be derived as
\begin{equation}
\begin{aligned}
 U(x', y', z')  = \mathcal{F} ^{-1}\left\{A'_0(f_x, f_y)\right\}H(f_x,f_y,z'), 
\end{aligned}
\label{e:fourier_shift}
\end{equation}
where 
\begin{equation}
\begin{aligned}
 A'_0(f_x, f_y) &= \mathcal{F} \left\{U(x,y,0)\right\} \exp{\left[i2\pi(f_x s_x + f_y s_y)\right]}.
\end{aligned}
\label{e:A'_0}
\end{equation}
Apparently, the Fourier transform in the shifted coordinate can be perceived as the same function in the original coordinate system multiplied by a phase shift proportional to the spatial displacement. 

\subsubsection{Self-consistent equation for round-trip transmission}
The situation for the off-axis receiver shares the same methodology with the tilted receiver. The round-trip transmission simulation can be disintegrated into several simulations of the field distribution passing through each optical element in the system. As shown in Fig.~\ref{f:offaxis-arch}, the simulations of light traveling through mirrors, lenses, the gain medium, and the free space between them are the same as the calculation we derived for the tilted circumstance. Yet the only difference is the stage between the gain medium and L2 since they are off-axis relative to each other. We define propagation operators $\mathcal{M}$ and $\mathcal{M}'$ representing the beam transmission from gain medium to L2 and from L2 to gain medium in the following:
\begin{equation}
\begin{aligned}
 \mathcal{M}[U(x,y)] = & \mathcal{F}^{-1}\{\mathcal{F} \left\{U(x,y)\right\} H(f_x,f_y,s_z+f) \\&\times\exp{\left[i2\pi(f_x \Delta x)\right]}\}.
\end{aligned}
\label{e:off}
\end{equation}
\begin{equation}
\begin{aligned}
 \mathcal{M}'[U(x,y)] = & \mathcal{F}^{-1}\{\mathcal{F} \left\{U(x,y)\right\} H(f_x,f_y,s_z+f) \\&\times\exp{\left[i2\pi(-f_x \Delta x)\right]}\}.
\end{aligned}
\label{e:off2}
\end{equation}

Thus, the self-consistent equation can be built by referring to Eq.~\eqref{e:self-consistent} as
\begin{equation}
    U(x,y,0) = \mathcal{CBDAM'BCCBMADBC}[U(x,y,0)].
    \label{e:self2}
\end{equation}

Stated that for mobility analysis of RB-SLIPT under NLOS propagation as shown in Fig. {f:sysandprob} with the receiver at both tilt and off-axis states, simulation process of beam transmission to tilt receiver is at first conducted, and then the off-axis situation is simulated.

\subsection{Eigenmode and Diffraction Loss}
With the Fox-Li method, we can obtain the self-reproducing mode, i.e., eigenmode, by iterating the self-consistent equations with all the above simulation cases. Suppose the self-reproducing mode $u(x,y)$ 
is formed at $t$-th iteration with the iterative FoX-Li algorithm, transmission factor $\gamma$, also the eigenvalue of the self-consistent equation, is defined as~\cite{siegmanlaser}
\begin{equation}
    \gamma = \frac{||u_{t+1}(x,y)||}{||u_t(x,y)||},
\end{equation}
where $||\cdot||$ represents the $\ell_1$ norm, leading to the calculation of the sum of the absolute values of each element in the vector. As illustrated in laser cavity characteristics, transmission factor indicates the transmission loss as loss$=1-|\gamma|^2$.

\subsection{Achievable Data Rate and Charging Power}
After obtaining the accurate transmission channel loss of RB-SLIPT, we can calculate the output laser power at the receiver with a linear relationship to input power $P_{\rm in}$ as~\cite{hodgson2005laser}
\begin{equation}\left\{
	\begin{aligned}
	&P_{\rm out} = \eta_{\rm extr}(\eta_{\rm excit}P_{\rm in}-C)\\
	&\eta_{\rm extr} =\frac{\eta_{\rm b}(1-R)\gamma}{1-R\gamma^2+\sqrt{R}\gamma[1/(\gamma_{S}\gamma)-\gamma_{\rm S}]} \\
	&C=A_{\rm g}I_{\rm S}\mid\ln{(\sqrt{R}\gamma_{\rm S}\gamma)}\mid
	\end{aligned},\right.
	\label{e:pout-methord}
	\end{equation}
where $\eta_{\rm excit}$, $\eta_{\rm b}$, $A_{\rm g}$, $I_{\rm S}$ are the excitation efficiency, overlap efficiency, cross-sectional area, and saturation intensity of the gain medium; $\gamma_{\rm S}$ is the internal loss generated in the gain medium; $R$ is the reflectivity of M2 at the receiver. Adopting the power splitting scheme as in \cite{mobility2}, we choose $\mu$ as the splitting factor and $P_{\rm out, IT}=\mu P_{\rm out}$ indicates the beam power for data transfer and $P_{\rm out, PT}=(1-\mu) P_{\rm out}$ for wireless charging, respectively.

For communications, intensity modulation is adopted, and the MTC of RB-SLIPT is a free-space optical intensity channel. Without restrictions on limited energy resources and safety issues, we assume the ratio of average received power to peak signal power $P_{\rm out, IT}$ lies in the range of $[0.5, 1]$~\cite{lowbound}. Then, the achievable data rate of RB-SLIPT, referring to the lower-bounded channel capacity, can be expressed as~\cite{datarate,totalNoise}
\begin{equation}\left\{
	\begin{aligned}
	&R_{\mathrm{b}}=\frac{1}{2} \log _{2}\left\{1+\frac{\left(\rho_{\rm pd} P_{\mathrm{out}, \mathrm{IT}}\right)^{2}}{2 \pi e \sigma_{\mathrm{n}}^{2}}\right\}\\
	&\sigma_{\mathrm{n}}^{2}=2 q\left(\rho_{\rm pd} P_{\mathrm{out}, \mathrm{IT}}+I_{\mathrm{bk}}\right) B+\frac{4 k T B}{R_{\mathrm{IL}}}
	\end{aligned},\right.
	\label{e:pout-methord}
	\end{equation}
where $\rho_{\rm pd}$ is the responsivity of PD, $e$ is the nature constant, and $\sigma_{\rm n}^2$ is the total noise variance. $q$ is the electron charge, $I_{\rm bk} = 5100 \mu $A is the photon current induced by background
radiation, $B=800$MHz is the bandwidth, $k$ is Boltzmann constant, $T$ is the temperature in
Kelvin, and $R_{\rm IL}$ is the load resistance of the PD.

For charging, we adopt an equivalent circuit model of PV panel, where the conversion from beam power incidents on it to charging power $P_{\rm chg}$ can be depicted as the following equations
\begin{equation}
    \left\{\begin{array}{l}
I_{\text {chg }}=\rho_{\rm pv}P_{\rm out,PT}-I_{\mathrm{d}}-\frac{V_{\mathrm{d}}}{R_{\mathrm{sh}}} \\
I_{\mathrm{d}}=I_{0}\left[\exp \left(\frac{qV_{\mathrm{d}}}{n_{\mathrm{s}} nkT}\right)-1\right] \\
V_{\mathrm{d}}=I_{\text {chg }}\left(R_{\mathrm{PL}}+R_{\mathrm{s}}\right) \\
P_{\text {chg }}=I_{\text {chg }}^2R_{\mathrm{PL}}
\end{array}\right.,
\end{equation}
where $\rho_{\rm pv}$ is the responsivity of PV, $I_{\rm d}$ and $V_{\rm d}$ are current and voltage passing through and on the diode, $I_0$ is the reverse saturation current, $n$ is the diode ideality factor, and $n_s$ is the number of series-connected cells. $R_{\rm PL}$ is the load resistance, and $R_{\rm sh}$ and $R_{\rm s}$ are shunt resistance and series resistance in the equivalent circuit model. 

\section{Implementation for Analytic Models}
For the implementation of the proposed analytical models, we should at first discretize the functions and denote the transmission process as the element-wise product of matrices. Thus, the sampling number requirements are essential for implementation as it determines the computational time and storage. In this section, we summarize the sampling requirements imposed by the receiver's moving: i) shifting along or tilting on x-axis or y-axis; 2) long-range moving along z-axis. Then, we present a specific design for implementing simulation for long-range transmission with an off-axis plane, which can reduce the required matrix dimensions and achieve feasibility. 
\subsection{Sampling Requirements}
\subsubsection{Requirements Imposed by Shift and Tilt}
According to~\cite{nonuniform1}, to eliminate the aliasing introduced by the free space transfer function sampling, the angular spectrum $A(f_x, f_y)$ needs to be cut off with an ideal low-pass filter based on the band-limited angular spectrum method. Suppose the cutoff frequency is $f_{\rm limit}$, the spatial frequency needs to satisfy

\begin{equation}
-f_{\rm limit} < f_x < f_{\rm limit},\quad -f_{\rm limit} < f_y < f_{\rm limit}
\label{e:freq_limit}
\end{equation}
$f_{limit}$ is determined with Nyquist's theorem:
\begin{equation}
f_{\rm limit} = \frac{1}{\lambda[(2x\Delta f)^2 + 1] ^ {1/2}}
\label{e:nyquist}
\end{equation}
where $z$ is the distance between the planes, $\Delta f$ is the frequency sampling interval. 

\begin{figure}[h]
    \centering
    \includegraphics[width=3.5in]{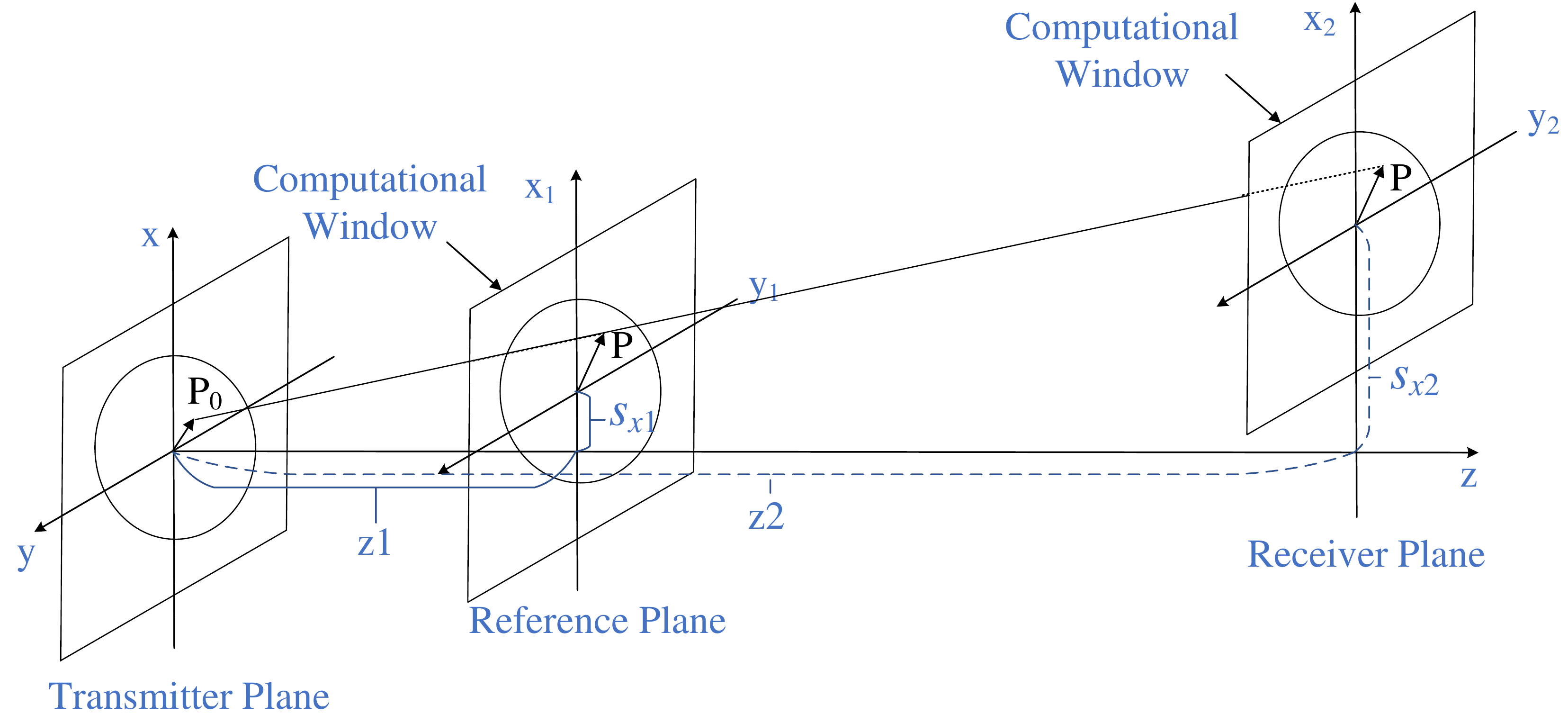}
    \caption{Off-axis receiver with long-range transmission.}
    \label{f:off-axis-long}
\end{figure}

Besides, sampling the input beam field spectrum is essentially duplicating the computational window due to the periodicity of FFT. A large tilting angle or shifting distance will result in a spectrum shift to the neighboring windows, leading to aliasing. Hence, a sufficiently large computational window is needed to make sure no aliasing is introduced. Taking the spatial frequency on x-axis to exemplify, the frequency after tilting should satisfy
\begin{equation}
-f_{\rm M} \leq \hat{f}_x = \cos(\varphi)f_x - \sin(\varphi)\sqrt{(1/\lambda)^2 - f_x^2} \leq f_{\rm M}
\label{e:tilt_freq}
\end{equation}
where $f_{\rm M} = 1/(2\Delta x)$. The relation between sampling interval and the tilting angle can be obtained as
\begin{equation}
\begin{aligned}
&\Delta x \leq \frac{1}{2|\cos(\varphi)f_x - \sin(\varphi)f_z|}\\
&\text{for}\ f_x \in \left[-f_{\rm limit} , f_{\rm limit}\right]
\end{aligned}
\end{equation}
where $f_z = \sqrt{(1/\lambda)^2 - f_x^2}$. For the $90^\circ$ tilting angle, i.e., the receiver is perpendicular to the transmitter, the sampling interval should satisfy
\begin{equation}
\begin{aligned}
\frac{1}{2|\cos(\varphi)f_x - \sin(\varphi)f_z|}&\approx \frac{1}{2f_z}\approx\frac{1}{2}\lambda,\\
f_z = \sqrt{(1/\lambda)^2 - f_x^2}&\approx 1/\lambda
\end{aligned}
\end{equation}

\subsubsection{Requirements Imposed by z-Axis Moving}
To comply with the transmission characteristics and application scenarios of RBS, a simulation with long-range (in terms of meters) and large angles (greater than 10$^\circ$) is needed. Under such circumstances, the plane of the gain medium on the transmitter's side is remote from the plane where the receiver's lens resides. Furthermore, the off-axis distance of the receiver will decide the angle between the transceivers' planes. As shown in Fig.~\ref{f:off-axis-long}, assume there is a certain angle between the transceivers, the distance $z_1$ between them is equivalent to an off-axis shift $s_{x1}$ on the receiver's side. By increasing the distance to $z_2$, the off-axis shift will be enlarged to $s_{x2}$ proportionally. As the distance continues to expand, however, the off-axis shift will exceed the margin of the computational window. Consequently, the optical field distribution can hardly be simulated on the receiver's side, whereas the next period of the optical field will be received due to the periodicity of FFT.

\begin{figure*}[h]
    \centering
    \includegraphics[width=6.1in]{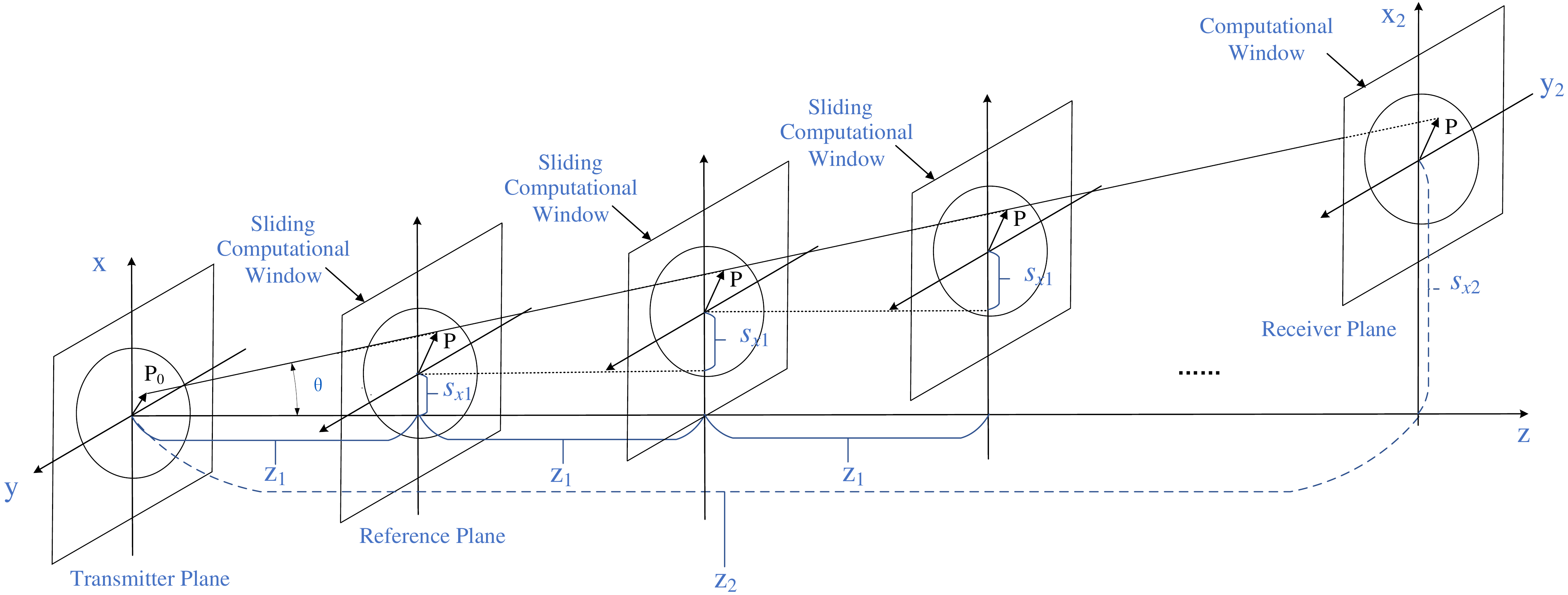}
    \caption{Implementation of long-range transmission with multi-hop sliding computational window design.}
    \label{f:seg-trans}
\end{figure*}

This problem can be alleviated to some extent by increasing the computational window. Nevertheless, the sampling number will simultaneously grow up significantly considering the sampling accuracy, leading to higher computational expense or even exorbitant memory consumption. To address the restrictive relation between the transmission distance and the size of the computational window, we propose a sliding computational window design to achieve the long-range mobility analysis without increasing the computational window.

\subsection{Design of Multi-Hop Sliding Computational Window}
The scheme of the sliding computational window is illustrated in Fig.~\ref{f:seg-trans}. Assume the distance between transceivers is $z_2$, the off-axis shift $s_{x2}$ of the receiver plane grows beyond the margin of the computational window for the transmitter plane. We can separate $z_2$ into several shorter distances $z_1$, and divide the free space transfer process into a series of transmission between the intermediate planes. With the same translation angle, the first intermediate plane has an off-axis shift $s_{x1} = s_{x2}z_1/z_2$ with respect to the transmitter's plane, and the second intermediate plane has the same shift relative to the first intermediate plane, and so on. By judiciously selecting $z_1$, we can make sure that $s_{x1}$ falls within the computational window. Thus, the memory insufficiency and exorbitant computational requirements can be eliminated by confining the size of the window by introducing more intermediate planes.

The relation between the size of computation window, $z_1$, $z_2$, $s_{x1}$ and $s_{x2}$ can be deduced as follows. Denote the radius of the cat's eye retroreflector as $r$, the size of the window is $2wr$ with the expansion factor is set to be $w$. We further assume the included angle between z-axis and the line linking transceivers is $\theta$, then the relation between them can be represented mathematically as:
\begin{equation}
\begin{aligned}
s_{x2} &= z_2\tan\theta\\
z_1 &= z_2s_{x1}/s_{x2}\\
s_{x1} &\leq wr
\end{aligned}
\label{e:s_z_relation}
\end{equation}

Since the radius $r$ and the distance $s_z$ are known, given the angle $\theta$, we can obtain the requirement for $z_1$ as:
\begin{equation}
z_1\leq \frac{wr}{\tan\theta}
\label{e:z1}
\end{equation}

Moreover, $z1$ should meet the sampling accuracy requirement for the diffraction transmission between two planes~\cite{FFT}.

For the situation where the receiver is tilted, a feasible technique is to set the intermediate planes to be tilted, i.e., the intermediate planes are parallel to the receiver. With this configuration, the coordinate system mapping for the tilted plane transmission will only be deployed once in the calculation of the propagation between the transmitter and the first intermediate plane. For the rest of the planes, free space transfer between  parallel planes will work. Specifically, the propagation distance between two planes is \begin{equation}
\Delta d =\frac{z_1}{\cos\theta}
\end{equation}

\section{Numerical Results and Discussions}
In this section, we will specifically analyze how the tilted angle of the receiver influences the transmission channel, which is reflected by the transmission factor with different system parameters. Besides, we show the computational time for conducting one round-trip transmission and the accuracy of different methods. We also conduct experiments with an off-axis receiver. To depict the energy/data transfer performance of RB-SLIPT under NLOS propagation, we plot the achievable data rate as well as charging power with different reflector positions and receiver's attitude angles.

\begin{table}[!htbp]
\centering
\caption{~Parameter of Energy/Data Transfer{\color{black}~\cite{PVpara,Pbck}}}
\vspace{.7em}
\begin{tabular}{ccc}
\hline
\textbf{Parameter}&\textbf{Symbol}&\textbf{Value}\\
\hline
\text{PV conversion responsivity}&$\rho_{\rm pv}$&{\color{black}$0.0161$A/W}\\
\text{Reverse saturation current}&$I_\text{0}$&{\color{black}$9.89\times 10^{-9}$A}\\
\text{Diode ideality factor}&$n$ & {\color{black}$1.105$}\\
\text{Number of serial PV cells}&$n_{\text{s}}$ & {\color{black}$40$}\\
\text{Series resistance}&$R_{\text{s}}$ & {\color{black}$0.93\Omega$} \\
\text{Shunt resistance}&$R_{\text{sh}}$ & {\color{black}$52.6$k$\Omega$}\\
\text{PV load resistor}&$R_{\text{PL}}$ & {\color{black}$100\Omega$} \\
\text{Temperature in Kelvin}&$T$ & {\color{black}$298.15$K}\\
\text{PD conversion responsivity}&$\rho_{\rm pd}$&{\color{black}$0.4$A/W}\\
\text{APD load resistor}&$R_{\text{IL}}$&$10$K$\Omega$~\cite{totalNoise}\\
\text{Splitting factor}&$\mu$&$0.01$\\
\hline
\label{t:SWIPTPara}
\end{tabular}
\end{table}

The numerical analysis of the RBS is conducted in MATLAB. The radius cat's eye front surface is defined as $r_c$, and the radius of the gain medium is $r_g$. Fast Fourier transform (FFT) and Fox-Li algorithm are used for calculating the resonant beam field distribution (i.e., self-reproducing mode). Then, appropriate sampling and zero-padding are required to avoid the aliasing in the FFT~\cite{FFT,FoxLi}. Generally, the sampling number is a power of $2$ and should meet the sampling requirements for the chirp function in the diffraction integral equation. Besides, to avoid aliasing, we also surround the aperture with zero elements with the window expanding factor. Finally, the Fox-Li iteration stops as the difference of the beam field distributions between two iterations is below $10^{-4}$. Moreover, the parameters for energy conversion and data transfer performance analysis are detailed in Table~\ref{t:SWIPTPara}.

\begin{figure*}[htpb]
\centering
\subfigure[]{
\includegraphics[width=7cm]{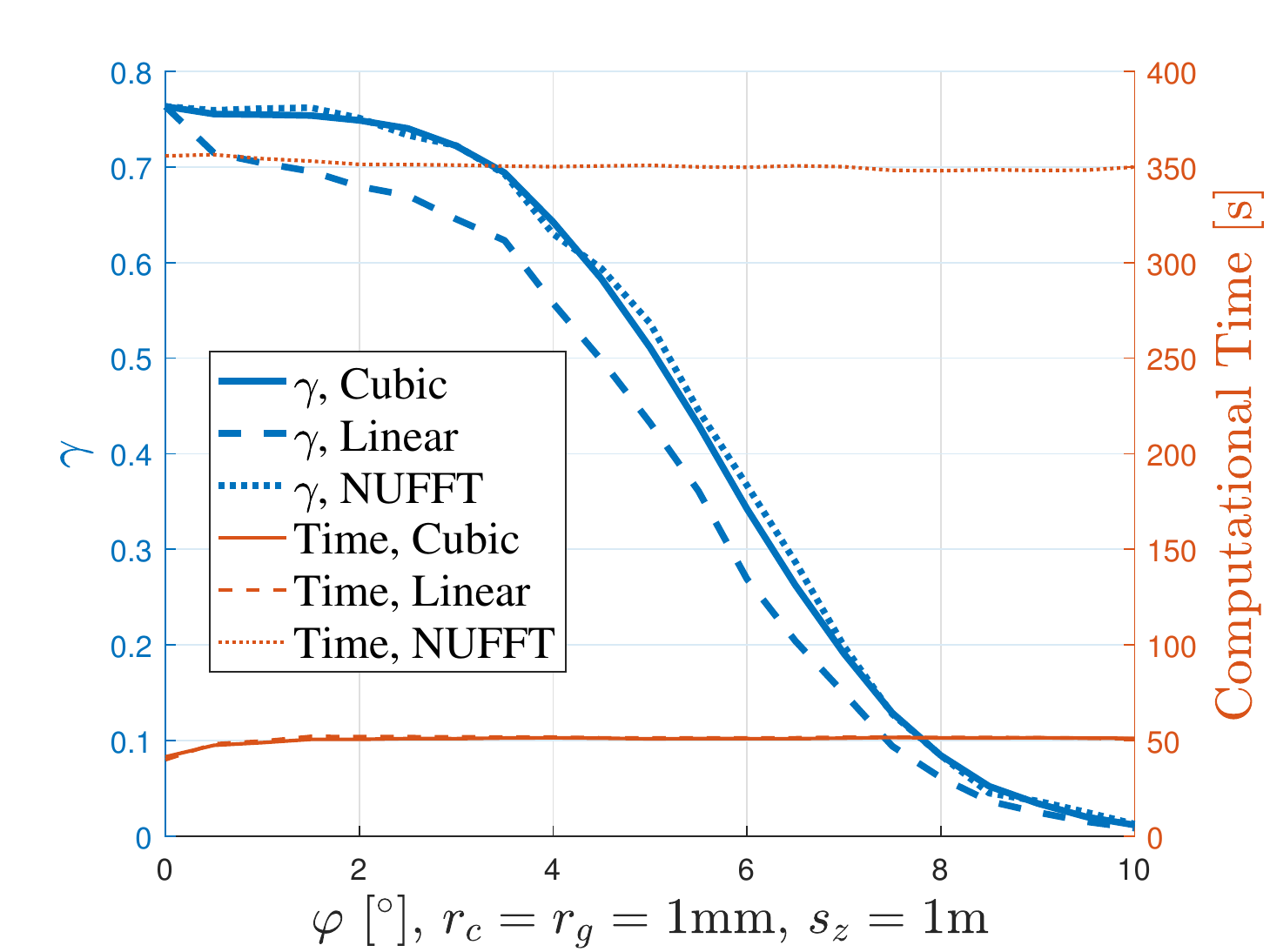}
}
\subfigure[]{
\includegraphics[width=7cm]{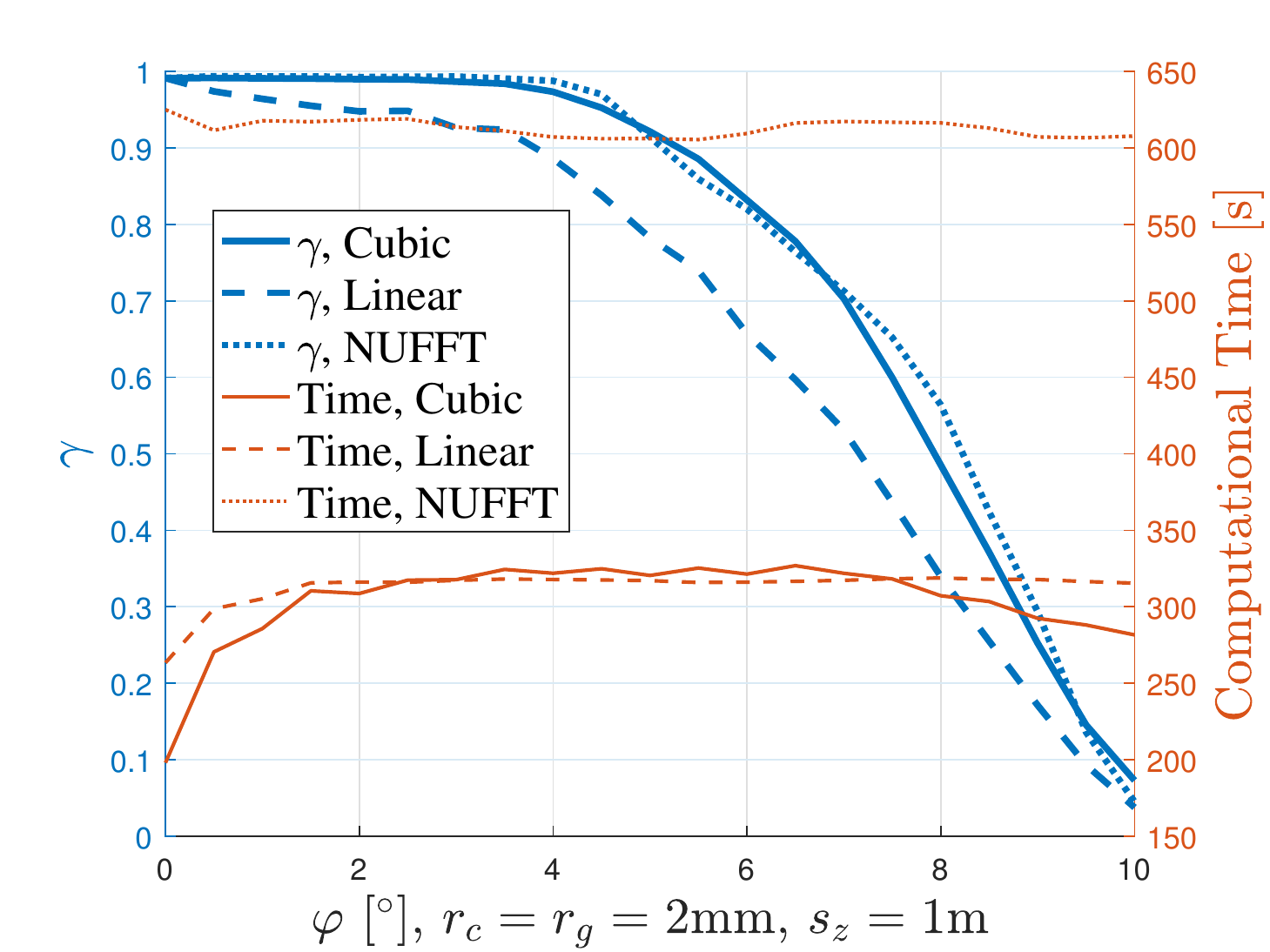}
}
\subfigure[]{
\includegraphics[width=7cm]{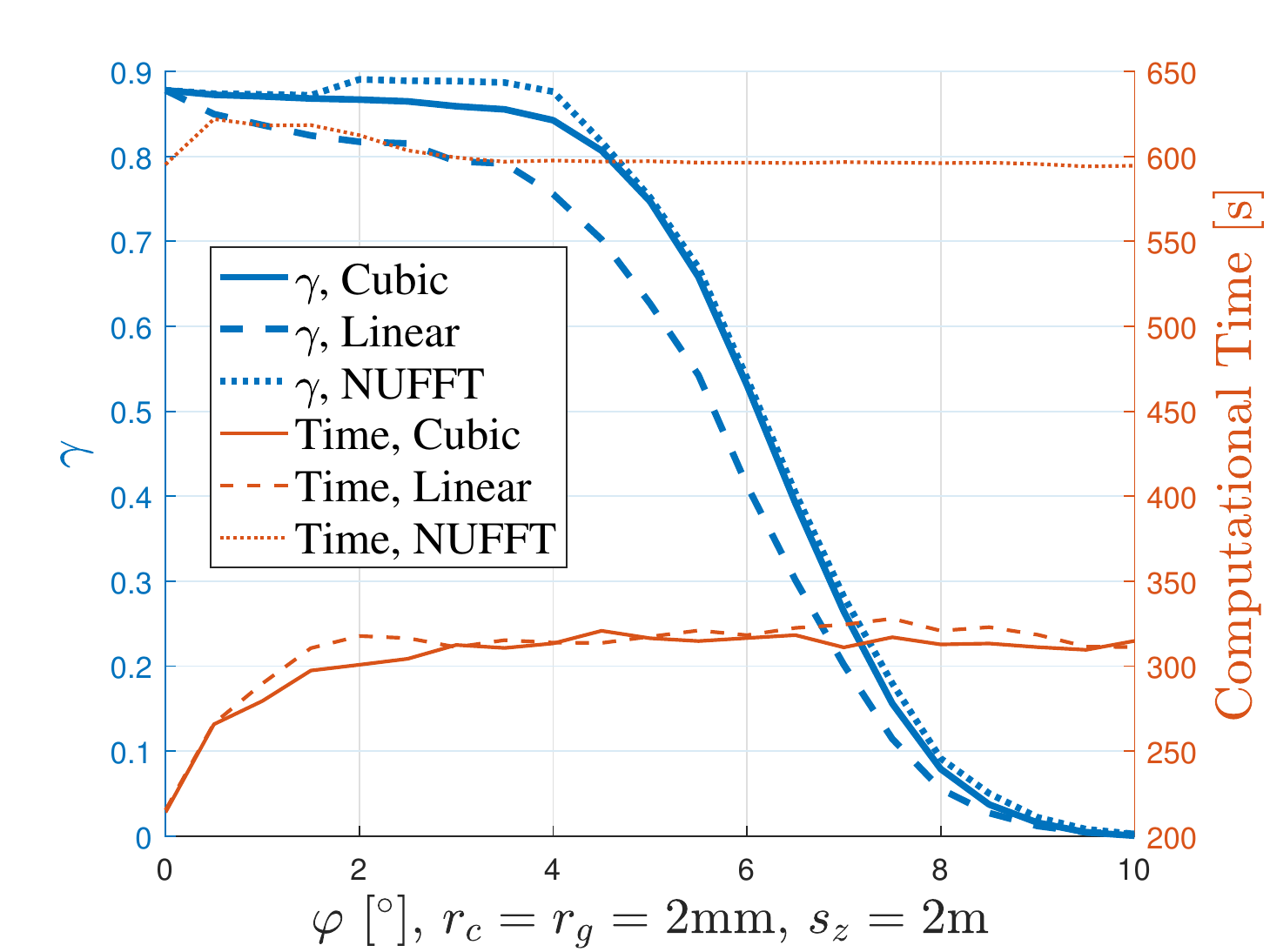}
}
\subfigure[]{
\includegraphics[width=7cm]{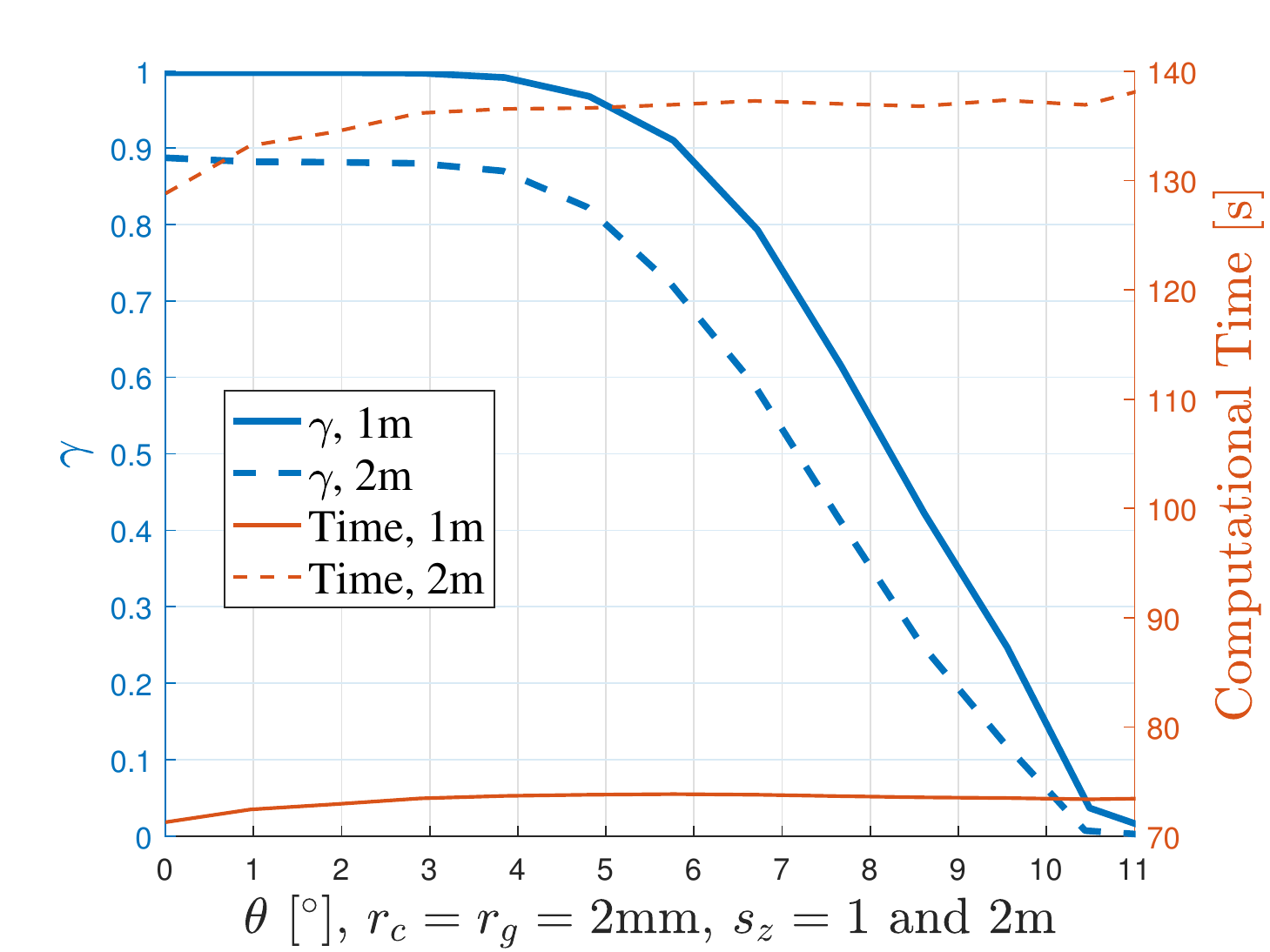}
}
\caption{Transmission factor $\gamma$ and computational time of one round-trip transmission as a function of tilted angle $\varphi$ and translation angle $\theta$ with 
(a) $r_c=r_g=1$mm and $s_z=1$m; (b) $r_c=r_g=2$mm and $s_z=2$m; (c) $r_c=r_g=2$mm and $s_z=1$m; and (d) $r_c=r_g=2$mm and $s_z=1$m and $2$m.}
\label{f:tiltshit}
\end{figure*}

\subsection{Receiver with Arbitrary Attitude Angle}
The computational process of optical wave propagation between large-size apertures faces challenges with large sampling numbers. On the one hand, a large sampling number means a significant dimension of a matrix, which requires large storage. On the other hand, computation for a large matrix will cost a lot of time. Both of the above issues will thwart the feasibility of the simulation. However, for analyzing the transmission channel characteristics of RBS, we should simulate the beam transmission between apertures with mm-level size over m-level distance. Thus, we first depict the transmission factor and computational time for one round-trip transmission as the radius of both gain medium and cat's eye are $1$mm while the z-axis distance is $1$m. Stated the focal length is $5$mm. 

Fig. \ref{f:tiltshit}(a) shows the changing trend with the increase of tilted angles. As illustrated above, there are three numerical methods simulating the beam propagation between tilted planes, i.e., NUFFT, cubic interpolation, and linear interpolation. For a cat's eye with the stated parameters, the maximum allowable tilt or translation angle is around $10^{\circ}$. As the tilted angle increases, the transmission factor decreases and reduces to nearly zero as the tilted angle is $10^{\circ}$. A larger transmission factor means smaller transmission loss and thus better system performance. Changing trend under the above three methods is similar, while the results under NUFFT and cubic interpolation are in good agreement. Meanwhile, the cubic and linear interpolation methods show better time-saving.

We change the parameters and conduct similar experiments as Figs.~\ref{f:tiltshit}(b) and (c). With $r_c=r_g=2$mm, the moving distance can be $2$m without the loss of transmission factor. In this case, we change the focal length of the cat's eye to $10$mm to keep the maximum allowable angle of the incident beam the same. A similar conclusion can be made as agreement with results under NUFFT and cubic interpolation and faster computational speed under interpolation methods. Moreover, for a cat's eye with the same parameters, a shorter z-axis moving distance results in larger transmission factors.

\subsection{Receiver with Arbitrary Spatial Position}
We also conduct experiments with an off-axis receiver to illustrate the design of a sliding computational window as in Fig.~\ref{f:tiltshit}(d). The transmission factor and computational time for one round-trip transmission are depicted with $r_c=r_g=2$mm and focus length $l=10$mm. As the translation angle increase, i.e., the off-axis distance $\Delta x$ increases, the transmission factors decrease under z-axis moving distances $1$m and $2$m. At the same time, they remain unchanged as the translation angle is below $4^{\circ}$. As mentioned above, the number of the sliding computational windows is determined by the z-axis moving length. Thus, Computational time under $s_z=1$m is significantly less than under $s_z=2$m as the number of time-costing computational processes between sliding computational windows are $34$ and $67$, respectively. Meanwhile, with the computational design of sliding windows, the required sampling number dropped by a factor of $34^2$ and $67^2$ under $1$m and $2$m transmission, respectively, in this simulation settings.

\begin{figure}[h]
    \centering
    \includegraphics[width=3in]{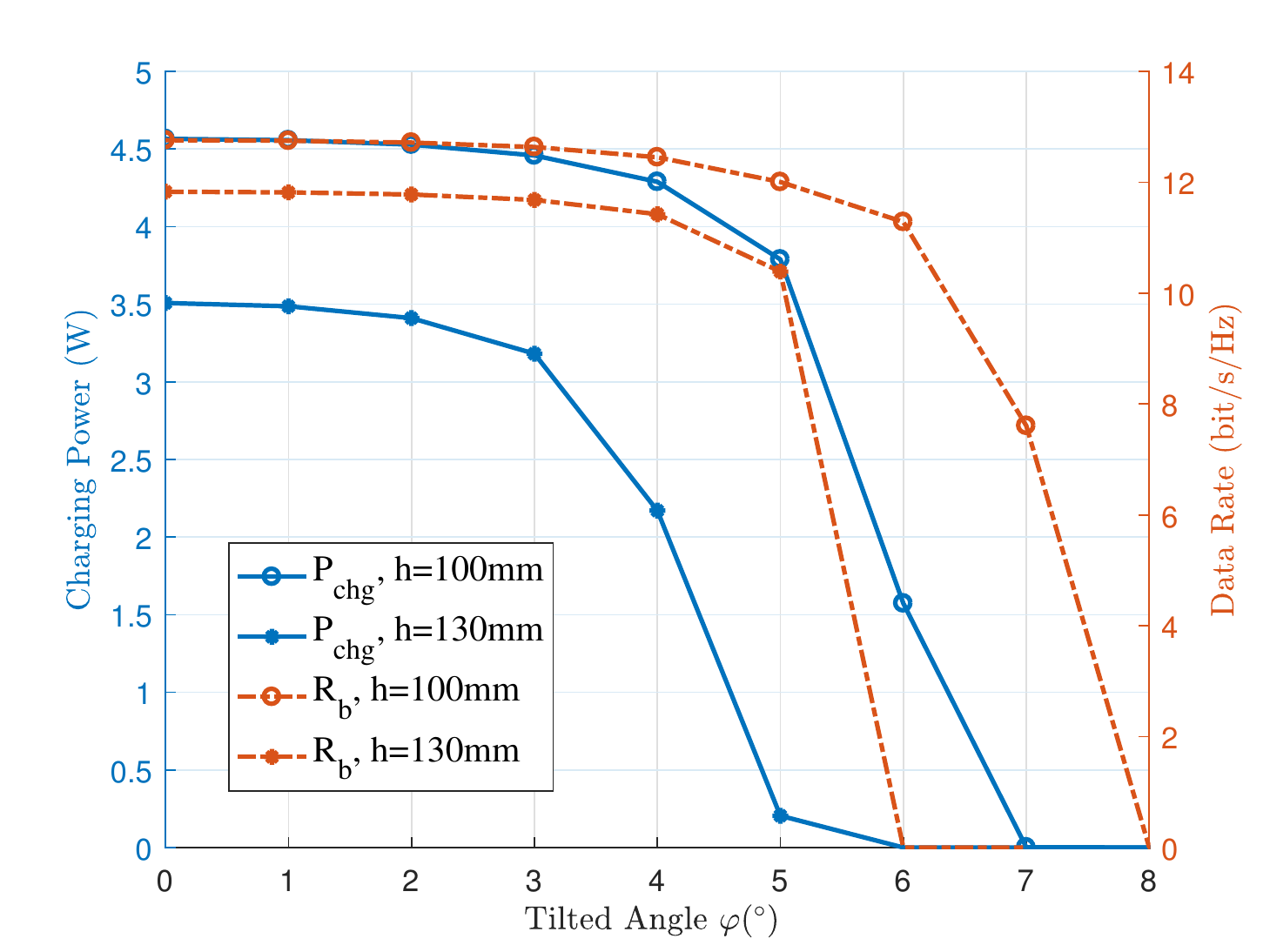}
    \caption{Charging power and achievable data rate of RB-SLIPT under NLOS propagation with different receiver tilted angles ($r_c=r_g=2$mm, $s_z=1$m).}
    \label{f:NLOS1}
\end{figure}

\begin{figure}[h]
    \centering
    \includegraphics[width=3in]{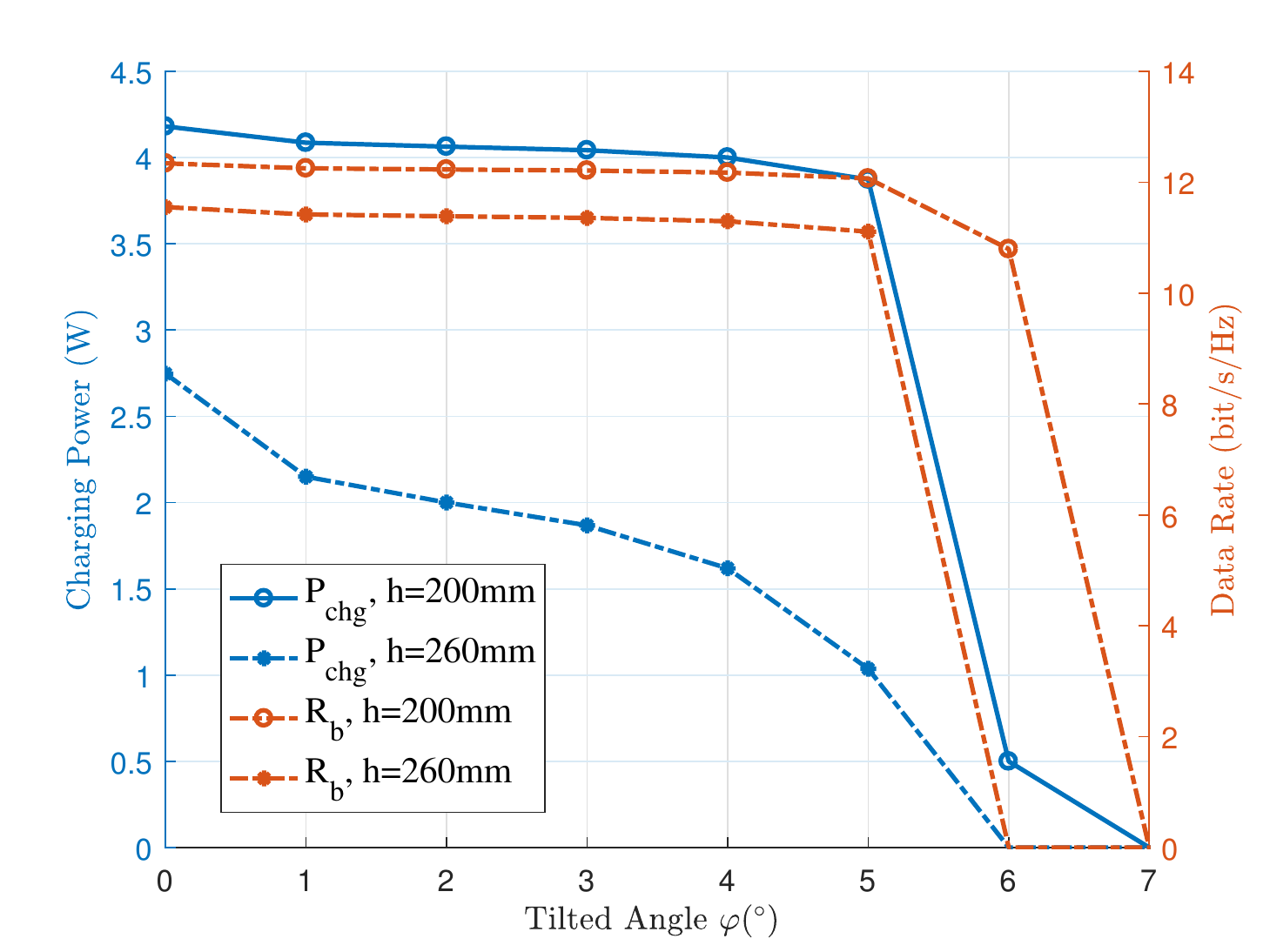}
    \caption{Charging power and achievable data rate of RB-SLIPT under NLOS propagation with different receiver tilted angles ($r_c=r_g=2$mm, $s_z=2$m)..}
    \label{f:NLOS2}
\end{figure}

\subsection{Energy/Data Transfer under NLOS Propagation}
Finally, we depict the energy/data transfer performance of RB-SLIPT under NLOS propagation with an arbitrarily placed receiver. As in Fig.~\ref{f:NLOS1}, we plot charging power $P_{\rm chg}$ as well as achievable data rate $R_{\rm b}$ as a function of the receiver's tilted angle $\varphi$ with different distances $h$ between the reflector and the original optical path. As the tilted angle grows with $h=100$mm and $s_z=1$m, both $P_{\rm chg}$ and $R_{\rm b}$ keep steady and drop down quickly as $\varphi$ reaches $5^{\circ}$. If $h=130mm$, both energy and data transfer performances reduce. Still, over $3.5$W charging power and $10$bit/s/Hz achievable data rate can be realized.
Similar experiments have been conducted with $s_z=2$m. With $h=200$mm, over $4$W charging power and nearly $12$bit/s/Hz achievable data rate can be realized as the receiver's tilted angle is smaller than $5^{\circ}$. With the increase of $h$, the system performance, especially the charging power, decreases obviously. That's because of the increase of transmission loss in NLOS scenarios with a further reflector. 

\subsection{Discussion}
We have analyzed the calculation results and computational time for the MTC analysis of RB-SLIPT using different numerical methods. Numerical results demonstrate that RB-SLIPT is capable of realizing $4$W charging power and $12$bit/s/Hz achievable data rate under NLOS propagation. To further improve the moving range of RB-SLIPT in NLOS scenarios, the reflector can be replaced by a retroreflector, i.e., a cat's eye retroreflector. Moreover, the intelligent reflective surface (IRS) can be adopted for futher enhancing the system performance. The analytical models for MTC with the above structures is worth further study. Moreover, some methods to speed up the computation should be further studied, i.e., methods for speeding up the Fox-Li process, methods for exploring other self-reproducing mode solving schemes, and methods for obtaining analytical solutions of transmission loss in RB-SLIPT. 

\section{Conclusions}
\label{sec:Conclusion}
In this paper, we proposed analytical models and simulation tools for MTC analysis of RB-SLIPT with a receiver in an arbitrary position and at an arbitrary attitude angle under NLOS propagation. We focused on the beam field propagation analysis with a tilted RB-SLIPT receiver and presented three numerical methods (i.e., NUFFT-based, cubic interpolation-based, and linear interpolation-based method) to implement the simulation. The transmission loss and accurate beam field profile can then be obtained for RB-SLIPT in NLOS scenarios. Next, we detailed the implementation of long-distance transmission with an off-axis RBS receiver by proposing a sliding computational window design, which can deal with the contradiction between long-range off-axis transmission and practically tolerable memory cost.  We used the transmission loss factor to indicate the feasibility of the proposed numerical methods and analyze the computational time. We also analyze the energy/data transfer performance of RB-SLIPT in NLOS scenarios, which demonstrates $4$W charging power and $12$bit/s/Hz data rate over $2$m distance.


\bibliographystyle{IEEEtran}
\bibliography{PositionCommunication_doubleCol}
\end{document}